\documentclass[preprint,showpacs,preprintnumbers,amsmath,amssymb,aps,nofootinbib]{revtex4}

\usepackage{graphicx}
\usepackage{dcolumn}
\usepackage{bm}

\newcommand{\BE}{\begin{equation}}
\def\EE{\end{equation}}
\def\BEA{\begin{eqnarray}}
\def\EEA{\end{eqnarray}}

\def\Dslash{\mathop{\not\!\! D}}

\begin{document}

\title{Methods for Pseudoscalar Flavour-Singlet Mesons with Staggered Fermions }

\author{Eric B. Gregory, Alan C. Irving, Chris M. Richards} 
\affiliation{Theoretical Physics Division, Department of Mathematical Sciences,
        University of Liverpool, Liverpool L69-7ZL, United Kingdom}

\author{Craig McNeile} 
\affiliation{Department of Physics and Astronomy, The Kelvin Building,
       University of Glasgow, Glasgow G12-8QQ, United Kingdom}

\date{\today}
    
\begin{abstract}
We present the first 2+1 flavour lattice QCD calculations 
of pseudoscalar flavour-singlet propagators using improved staggered fermions.
We explore the relevant techniques and discuss prospects for the larger
scale studies now in progress. The disconnected correlator is shown to
have a highly non-Gaussian distribution and reliable estimates of the
errors require care. 

\end{abstract}
   

\maketitle

\section{INTRODUCTION} \label{se:INTRODUCTION}

The high mass of the $\eta'$ meson relative to the 
light pseudoscalar mesons is thought to
be due to topology and the complexities of
the QCD vacuum \cite{Witten:1979vv, Veneziano:1979ec}.
The mass of the $\eta'$ meson is a clear experimental
signal of the complexities of the  QCD vacuum that are normally
obscured by confinement, so a first principles calculation
of the mass would be an important
milestone in taming non-perturbative QCD.
A robust first-principles calculation of the spectrum of the $\eta$/$\eta'$ 
system should help shed light on the workings of the fermion sea, and the 
mechanism by which quark loops elevate the mass of the singlet meson above 
that of the octet meson \cite{Witten:1979vv, Veneziano:1979ec}.

The computation of the mass of the $\eta'$ meson
has always been one of the goals of the program of hadron spectroscopy
from lattice QCD calculations.
There have been many 
lattice studies that have 
computed the mass spectrum of 
pseudoscalar flavour-singlet mesons with $N_f=2$ flavours
of fermions \cite{Itoh:1987iy,McNeile:2000hf,Struckmann:2000bt,Lesk:2002gd,
Schilling:2004kg,DeGrand:2002gm,Venkataraman:1997xi,Kogut:1998rh,
Fukaya:2004kp}.
Recently the JLQCD/CP-PACS collaboration reported on
a preliminary calculation of the $\eta$ and $\eta'$
mesons with 
$2+1$ flavours of Wilson fermions~\cite{Aoki:2006xk} .

Improved staggered fermions have provided some 
of the most impressive results in 
hadron spectroscopy
for flavour non-singlet quantities~\cite{Davies:2003ik,Aubin:2004wf}.
The next stage is to use improved staggered fermions 
to compute the mass spectrum of 
light singlet pseudoscalar mesons.
Improved staggered fermions have the 
advantage that they are fast to simulate and allow
lattice QCD calculations with
high statistics at relatively light quark masses, both 
requirements for accurate calculation of 
the masses of singlet pseudoscalar mesons.

However, lattice QCD calculations of the masses of the 
$\eta$ and  $\eta'$ mesons face several challenges. 
The lattice QCD calculations of the flavour singlet 
mesons involve disconnected correlators which are computationally
more expensive to compute than connected
correlators. Also
disconnected diagrams are inherently noisy in lattice simulations, so
high statistics are required. The disconnected diagram
for the pseudoscalar mesons is related to the
topological charge of the gauge configuration which in some cases has been 
seen to have  longer autocorrelation times than other quantities in 
lattice QCD simulations. To be sure statistical fluctuations and 
any autocorrelation times are well under control
high statistics are required for an accurate lattice QCD calculation.

Although lattice QCD calculations that use improved staggered 
fermions have been successfully tested against experiment
for many quantities~\cite{Davies:2003ik,Aubin:2004wf},
the formalism has not been proved to be fully 
correct because of the ``fourth-root trick'' employed to achieve the 
required flavour structure.
There is increasing theoretical
work~\cite{Durr:2005ax,Sharpe:2006re} 
that suggests that there are no problems
with the improved staggered 
fermion formalism in the continuum limit,
however the issue has not been completely settled.
The $\eta\eta^\prime$ system is a good place to
check the validity of the rooting of the sea quark determinant,
because of the important role the sea quark loops play
in raising the mass of the $\eta^\prime$ meson. Creutz
has recently claimed that the mass
of the $\eta'$ meson is a place where the improved
staggered fermion formalism 
may not work~\cite{Creutz:2007yg,Creutz:2007rk}.
Kilcup and Venkataraman~\cite{Venkataraman:1997xi}
have studied the flavour singlet pseudoscalar meson
with naive staggered fermions in unquenched QCD.

The $\eta$ and $\eta'$ mesons can both be studied
with lattice QCD calculations 
that include the dynamics of $2+1$ flavours of sea quarks.
In a quark model 
treatment~\cite{Donoghue:1992dd} of the $\eta$ and $\eta'$ mesons
in terms of light and strange quarks:
\begin{eqnarray}
\eta & \approx &  0.58 ( \overline{u} \gamma_5 u  + \overline{d} \gamma_5 d ) - 0.57 \; \overline{s} \gamma_5 s 
\nonumber \\
\eta' & \approx & 0.40 ( \overline{u} \gamma_5 u  + \overline{d} \gamma_5 d ) + 0.82 \; \overline{s} \gamma_5 s\ .
\nonumber
\end{eqnarray}
This suggests that both the $\eta$ and $\eta^\prime$ mesons
contain mixtures of light and strange quarks. The
$\overline{u}u  + \overline{d}d$ and
$\overline{s} s$ interpolating operators will both
couple to the $\eta$ and $\eta^\prime$ mesons; in the 
singlet pseudoscalar channel the lightest state
will be the $\eta$ and the first excited state 
the $\eta^\prime$ meson. This is an additional complication
of lattice QCD calculations that include 2+1 flavours
of sea quarks over those which only include
2 flavours of sea quarks. In $N_f = 2$ full lattice
QCD calculations the emphasis is on finding the 
large mass splitting between the singlet pseudoscalar
meson and the flavour non-singlet flavour mesons.
With 2+1 flavours of sea quarks it also interesting to
see how the mass of the $\eta$ meson is reproduced
because this is sensitive to the mixing of
$\overline{q}\gamma_5 q$ and $\overline{s}\gamma_5 s$ loops.
The reader is referred to the review by Feldmann~\cite{Feldmann:2002kz}
for a recent discussion of $\eta\eta'$ mixing.
The modern way to describe $\eta\eta'$ mixing 
is via leptonic decay constants~\cite{Feldmann:2002kz}.
See~\cite{McNeile:2000hf} for an attempt to compute
the relevant decay constants in $N_f=2$ unquenched
QCD.

In this paper we report on the lattice methods required
to study the singlet pseudoscalar mesons using improved
staggered fermions. This work is the necessary starting point
for a project that uses a large number of $N_f=2+1$ configurations.
For some cross-checks on the unquenched calculations we
also study the singlet pseudoscalar meson in quenched
QCD and briefly investigate the role of topology.

The general plan of the paper is as follows:
In Section~\ref{se:theory} we describe the theory behind
computing the correlators of singlet pseudoscalar mesons using improved
staggered fermions. Next, in Section~\ref{sim_meas}, we describe
the details of the calculation and report on the algorithmic work
to compute the required disconnected diagrams. Section~\ref{se:results} contains the results from the 
quenched and unquenched calculations and in 
Section~\ref{disc_stat_section} we discuss the statistics
of singlet and non-singlet correlators. After that we estimate the 
number of configurations required to get a given accuracy in 
Section~\ref{se:numberneeded}. The paper ends with our conclusions
in Section~\ref{se:conclusions}.

\section{Theoretical background}   \label{se:theory}

For notation, we use SP for the singlet pseudoscalar meson and
NP for the non-singlet pseudoscalar meson. The additional complications
due to staggered fermions are discussed later in this section.
We begin with the expression of the pseudoscalar singlet propagator for $N_f$
degenerate flavours of fermions:
\begin{equation}
G_{SP}(x',x)=\langle\sum_{i=1}^{N_f}\overline{q}_i(x')\gamma_5q_i(x')\sum_{j=1}^{N_f}\overline{q}_j(x)\gamma_5q_j(x)\rangle.
\end{equation}

This 
propagator gives rise to two different types of diagrams. There are $N_f$
connected diagrams:
\begin{equation}
\langle\sum_{i}\overbrace{\overline{q}_i(x')\gamma_5\underbrace{q_i(x')\sum_{j}\overline{q}_j}(x)\gamma_5q_j}(x)\rangle,
\end{equation}
and $N_f^2$ disconnected terms:
\begin{equation}
\langle\sum_{i}\overbrace{\overline{q}_i(x')\gamma_5q_i}(x')\sum_{j}\overbrace{\overline{q}_j(x)\gamma_5q_j}(x)\rangle.
\end{equation}
The connected term in the $N_f=2$ flavour 
symmetric case is the same as that for the (non-singlet) pion 
propagator. 

So we can write:
\begin{equation}
\label{degenerate_singlet_prop}
G_{SP}(x',x)= N_fC(x',x)-N_f^2D(x',x),
\end{equation}
where the extra fermion loop in the disconnected diagram gives rise to 
the relative minus sign, and
\begin{equation}
G_{NP}(x',x)= N_fC(x',x).
\end{equation}

\subsection{$D/C$ ratio}

In full Euclidean QCD we expect that the pion 
propagator should decay exponentially at large time separations 
as
\begin{equation}
G_{NP}(t) = Ae^{-m_{NP} t}
\end{equation}
and similarly for the singlet propagator:
\begin{equation}
G_{SP}(t) = Be^{-m_{SP} t}\, .
\end{equation}
As noted above the singlet propagator contains a disconnected and 
connected part, the latter being proportional to the pion propagator:
\begin{equation}
\label{singlet_prop}
G_{SP}(t) = Ae^{-m_{NP} t}-N_f^2D(t)\, .
\end{equation}

Taking the ratio $R(t)$ of the disconnected to connected parts at large 
time-separation $\Delta t$ then suggests
\begin{equation}
\label{degenerate_ratio_eq}
R(t)=\frac{N_f^2D(t)}{N_fC(t)}= 1-\frac{B}{A}e^{-(m_{SP}-m_{NP}) t}\, .
\end{equation}
This derivation of (\ref{degenerate_ratio_eq}) 
requires one to assume the same action is
governing both sea quarks and valence quarks --- in systems where the physics of the
valence and sea quarks differ this expression may no longer apply. An 
important case of this is in the quenched limit where the $D/C$ ratio 
is \cite{Venkataraman:1997xi}
\begin{equation}
\label{quenched_ratio_eq}
R(t)= A^\prime + B^\prime t\, .
\end{equation}

\begin{figure}[t]
\resizebox{5.0in}{!}{\includegraphics{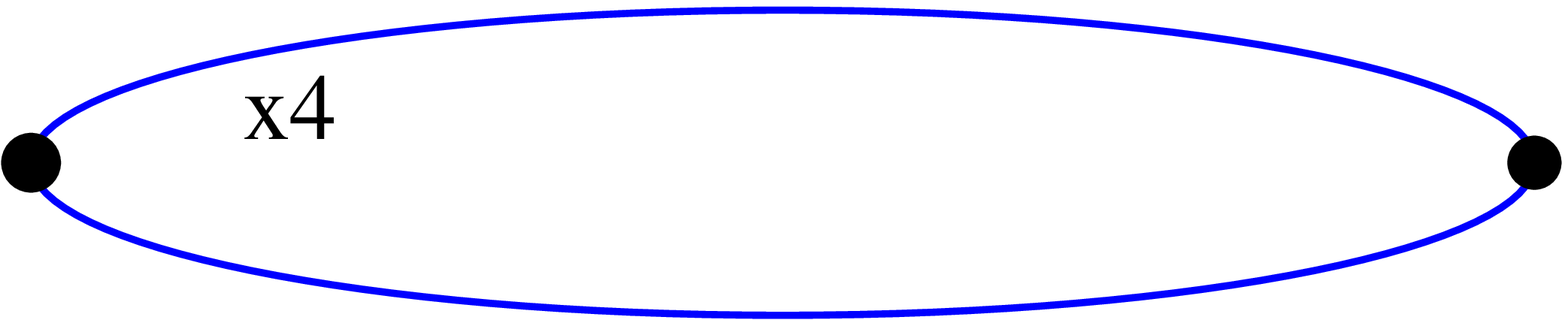}}
\resizebox{5.0in}{!}{\includegraphics{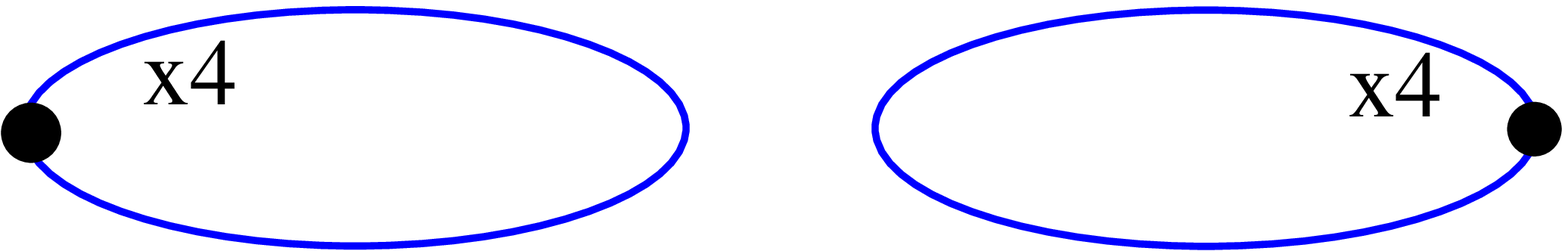}}
\caption{The valence contribution to the connected (top) and disconnected 
(bottom) diagrams. There are natively $N_v=4$ valence tastes in the 
staggered formulation, so each of these loops introduces a factor of four.
\label{loops_fig}}
\end{figure}

Another important case, relevant for staggered fermions, 
is when the number of flavours
$N_f$ of sea quarks differs from the native number of valence fermions, 
$N_v=4$. In practice, this is the case for most staggered simulations --- while
the fourth-root trick reduces four degenerate sea flavours to the desired 
$N_f$, there remain four native valence flavours, or tastes, which 
contribute to singlet propagators. 
In the connected contribution (top diagram in Fig. \ref{loops_fig}) a single
valence loop connecting the endpoints (meson interpolating operators)  
has four tastes of fermions circulating, 
introducing a factor of four. However in each of the terms of the 
disconnected contribution (bottom diagram in Fig. \ref{loops_fig}) there are 
two valence loops, one at each endpoint, each contributing a factor of four.
The $D/C$ ratio therefore naively has an extra factor of $N_v$ and, if 
the staggered formulation (incorporating a fourth-root of the determinant)
correctly reproduces $N_f$ sea flavours, we expect:
\begin{equation}
\label{ratio_eq_too_large}
R(t)=\frac{N_f^2D(t)}{N_fC(t)} = N_v(1-Ke^{-(m_{SP}-m_{NP}) t})\, .
\end{equation}
In the numerical work described in the following sections we implicitly 
rescale the disconnected contributions $D$ by a factor of $1/4$ so as to 
correct for the extra valence tastes, a procedure which Sharpe calls 
``valence rooting''~\cite{Sharpe:2006re} and was used  
by Venkataraman and Kilcup in~\cite{Venkataraman:1997xi}.
We will continue to refer to use the 
NP and SP notation, rather than introduce some new acronyms 
based on taste singlet notation.

There remains a possibility that the fourth-root trick could introduce 
some other pathologies at finite lattice spacing \cite{Creutz:2007yg}, 
e.g.~the wrong number of sea 
flavours or a mismatch between sea and valence fermion masses 
\cite{Hasenfratz:2005ri}. Such issues should, in principle, 
be detectable in the $D/C$ ratio.


For the case of $N_f$ degenerate flavours there is only
one possible interpolating operator for flavour singlet
mesons for non-degenerate flavours of sea quarks there are
more possible interpolating operators. In quark model
inspired discussions of the $\eta\eta^\prime$ 
mixing~\cite{Donoghue:1992dd},
the calculations use the SU(3) octet $\eta_8$ 
and singlet $\eta_0$
basis states
\begin{eqnarray}
\eta_0 & = & \frac{\overline{u} \gamma_5 u 
+ \overline{d} \gamma_5 d + \overline{s} \gamma_5 s} {\sqrt{3}}
\nonumber \\
\eta_8 & = & \frac{\overline{u} \gamma_5 u 
+ \overline{d} \gamma_5 d - 2 \overline{s} \gamma_5s} 
 {\sqrt{6}}\ .
 \\
\end{eqnarray} 
These basis states mix, because the $SU(3)$  symmetry is broken by the large
mass of the strange quark, to form the physical $\eta$ and $\eta^\prime$ mesons.

Another possible basis for flavour singlet pseudoscalar
mesons is the quark basis,~\cite{Feldmann:2002kz}
\begin{eqnarray}
\label{quark_basis}
\eta_q & = & \frac{\overline{u} \gamma_5 u 
+ \overline{d} \gamma_5 d } {\sqrt{2}}
\nonumber \\
\eta_s & = & \overline{s} \gamma_5 s.
\end{eqnarray} 
The $\eta_q$ and $\eta_s$ will both also
couple to the $\eta$ and $\eta^\prime$ mesons,
and are valid interpolating operators for
flavour singlet pseudoscalar mesons.
Each of these interpolating operators ($\eta_q$, $\eta_s$,
$\eta_0$ and $\eta_8$) will have the $\eta$ meson as the ground state
and the $\eta^\prime$ as the first excited state.

In principle the $0^{-+}$ glueball could mix
with the $\eta$ and $\eta^\prime$ mesons. This would mean
that a $0^{-+}$ glueball interpolating operator should 
also be used to study $\eta$ and $\eta^\prime$ mesons.
However quenched glueball
simulations~\cite{Morningstar:1997ff,Morningstar:1999rf} suggest
the mass of the $0^{-+}$ glueball to be around $2.6$ GeV.  Since this
is far from the mass of the $\eta^\prime$, we do not consider the
$0^{-+}$ glueball interpolating operator in what follows.  Hart and
Teper~\cite{Hart:2001fp} studied the $0^{-+}$ glueball correlators in
unquenched QCD and, despite large statistical errors, claimed the results
 were similar to those from quenched QCD.

The $\eta$ does not decay via the strong interaction, 
hence the mass of the $\eta$ is a 
gold-plated quantity~\cite{Davies:2003ik} that should 
be possible to compute very accurately. The $\eta'$ 
decays via the strong interaction but it has a small
width of 0.2 MeV~\cite{Yao:2006px} (relative to that of
the $\rho$ meson, for example). 
The dominant strong interaction decay of the 
$\eta'$ is via the decay to $\eta \pi \pi$ and thus
we expect that this decay channel is not open for the masses
we use. Thence we expect that an accurate calculation
of the $\eta'$ mass should be possible in principle.

For our $N_f=2+1$ simulations, 
if we use the $\eta_0$ interpolating operator,
we must generalise
(\ref{degenerate_singlet_prop}) and (\ref{degenerate_ratio_eq}) to the 
case of non-degenerate flavours. The full propagator becomes
\begin{equation}
\label{nondegenerate_singlet_prop}
G_{SP \; SU3}(t)= 2C_{qq}(t)+C_{ss}(t)-4D_{qq}(t) + 4D_{qs}(t)+ D_{ss}(t),
\end{equation}
and likewise the $D/C$ ratio is
\begin{equation}
\label{nondegenerate_ratio_eq}
R(t)_{SU3}  =\frac{4D_{qq}(t) + 4D_{qs}(t)+ D_{ss}(t)}{2C_{qq}(t)+C_{ss}(t)},
\end{equation}
where $D_{qq}$, $D_{qs}$, and $D_{ss}$ represent the disconnected correlators
constructed with two light quark loops, one light and one strange quark loop, 
and two strange quark loops, respectively. Likewise $C_{qq}$ and $C_{ss}$
are the connected correlators measured with light and with strange quark masses,
respectively.

\subsection{Fitting the singlet propagator} \label{se:fittvary}

With $2+1$ flavours of light quarks we have the potential to investigate 
the spectrum of both the $\eta$ and the $\eta'$ states. 
As discussed in Section~\ref{se:INTRODUCTION}, in $2+1$
flavour calculations, the $\eta'$ meson is the first
excited state contributing to the pseudoscalar correlator.
One approach
would be to measure the appropriate connected and disconnected correlators,
assemble the full singlet propagator in 
(\ref{nondegenerate_singlet_prop}) and then fit to a multi-exponential form:
\begin{equation}
G(t)_{SP} =A_1e^{-m_{SP1} t} + A_2e^{-m_{SP2}t}.
\end{equation}
However, fitting a multi-exponential expression with several parameters
to data consisting of a single correlator can be difficult 
when that data is precise, 
and even more so when dealing with inherently noisy 
disconnected correlators.

In a multi-channel approach, treating the non-strange ($q$) and strange ($s$) contributions
separately, using the quark basis (\ref{quark_basis}) one would form the matrix
\begin{eqnarray}
\label{matrix_prop}
{\bf G}(\Delta t)&=&
 \left[ 
\begin{array}{cc}
\eta^\dagger_q(\Delta t)\eta_q(0) &\eta^\dagger_q(\Delta t)\eta_s(0)\\
\eta^\dagger_s(\Delta t)\eta_q(0) & \eta^\dagger_s(\Delta t)\eta_s(0)
\end{array} \right]\nonumber\\
&=&
 \left[ 
\begin{array}{cc}
{\bf C}_{qq}(\Delta t)-2{\bf D}_{qq}(\Delta t) &  -\sqrt{2}{\bf D}_{qs}(\Delta t) \\
-\sqrt{2}{\bf D}_{sq}(\Delta t) & {\bf C}_{ss}(\Delta t)-{\bf D}_{ss}(\Delta t)
\end{array} \right],
\end{eqnarray}
and then fit 
\begin{equation}
{\bf G}(t)= {\bf A^T } {\bf e^{- m t}} {\bf A}
\label{eq:varyFIT}
\end{equation}
where ${\bf A }$ is a matrix of amplitudes and 
${\bf e^{- m t}}$ is a diagonal matrix. 
This variational approach to fitting is one of the main methods of extracting
masses and decay constants of excited mesons
from lattice QCD calculations. The variational method
is reviewed in~\cite{McNeile:2000xx} and 
used 
in~\cite{Luscher:1990ck,Burch:2006cc,Burch:2006dg,McNeile:2006qy,Hart:2006ps}.
The recent work on the $\eta$ and $\eta^\prime$
mesons by the CP-PACS/JLQCD 
collaboration used the variational method~\cite{Aoki:2006xk}.

Further improvement might be obtained by replacing the ${\bf C}$s and 
${\bf D}$s in (\ref{matrix_prop}) by matrices of correlators formed 
with the various combinations of fuzzed and point source and sink operators.
We leave investigation of such variational 
fits for future work.

\section{Simulation and measurement}
\label{sim_meas}

\subsection{Configuration ensembles}

We performed singlet correlator measurements on the $N_f=0,2,$ and $2+1$
ensembles listed in Table \ref{ensembles}. We used $16^3\times32$ 
lattices for
algorithm tuning as discussed below, and $20^3\times 64$ lattices for physics measurements.
The latter are primarily the so-called ``coarse'' lattice gauge configurations 
generated by the MILC collaboration~\cite{Bernard:2001av} with the 
``Asqtad'' improved staggered 
action \cite{Orginos:1998ue,Orginos:1999cr,Orginos:1999kg,Lepage:1998vj}.

\begin{table}
\begin{tabular*}{0.85\textwidth}{@{\extracolsep{\fill}}|llllll|}
\hline
$N_f$           & $10/g^2$ & $L^3\times T$ & $am_{\rm sea}$ & $am_{\rm val}$ & $N_{\rm cfg}$ \\
\hline
\hline
0         &   8.0  & $16^3\times 32$  & ---  & 0.020 & 76 \\
2         &   7.2  & $16^3\times 32$  & 0.020 & 0.020 & 268 \\

\hline
\hline
0         &   8.00 & $20^3\times 64$  & ---  & 0.020 & 408 \\
0         &   8.00 & $20^3\times 64$  & ---  & 0.050 & 408 $\longrightarrow$ 6154\\
\hline
2         &   7.20 & $20^3\times 64$  & 0.020 & 0.020 & 547 \\
\hline
2+1       &   6.76 & $20^3\times 64$  & 0.007, 0.05 & 0.007, 0.05 & 422\\
2+1       &   6.76 & $20^3\times 64$  & 0.010, 0.05 & 0.010, 0.05 & 644\\
2+1       &   6.85 & $20^3\times 64$  & 0.05, 0.05 & 0.05, 0.05 & 369\\
\hline
\end{tabular*}
\caption{Ensembles used for singlet calculations. 
\label{ensembles}}
\end{table}

For analysis of the fluctuations in disconnected correlators 
(Section \ref{disc_stat_section}) we extended the $\beta=8.00$ quenched 
ensemble from 408 configurations to 6154 configurations. From 
the final configuration in the MILC ensemble, configuration
 4090, we 
initiated ten separate Markov streams with 10 sweeps, each consisting of
four over-relaxation steps and and a quasi-heatbath hit, between saved
configurations. These are listed in Table \ref{extended_quenched_configs}. 
\begin{table}
\begin{tabular*}{0.85\textwidth}{@{\extracolsep{\fill}}|l||ccccccccccc|c|}
\hline
Label & MILC & S0 & S1 & S2 & S3 & S4 & S5 & S6 & S7 & S8 & S9 & all \\
\hline
Configs & 408 & 2455 & 413 & 414 & 409 & 411 & 409 & 410 & 409 & 412 & 412 & 6154 \\
\hline
\end{tabular*}
\caption{Quenched configurations in the MILC and extended ensemble streams.
\label{extended_quenched_configs}}
\end{table}

\subsection{Connected and disconnected correlators}
The numerical calculation of the singlet propagator is performed
in two parts --- the calculation of the connected contributions, and the calculation of the
 disconnected contributions.
The former is relatively straightforward --- inversions on point source vectors 
produce quark propagators which are then multiplied together with an appropriate
meson operator to produce a meson propagator.

There are in principle two different meson operators which couple to the 
flavour-singlet pseudoscalar meson. 
These are the $(\gamma_5\otimes{\bf 1})$
and the $(\gamma_4\gamma_5\otimes{\bf 1})$, using the 
Kluberg-Stern~\cite{KlubergStern:1983dg} notation
for the staggered meson operators.
 The former has the
quark and anti-quark sources separated by four links, on opposite corners of the
hypercube, while the latter  has the quark and anti-quarks separated by three links, sited at opposite
corners of the {\em spatial} cube.
We write the operator as $(\gamma_5\otimes{\bf 1})$ to show that the meson 
has pseudoscalar Dirac structure, but is a singlet in Kogut-Susskind taste
space.

As is generally the case with
staggered meson operators, the operator 
which couples to the 
$(\gamma_4\gamma_5\otimes{\bf 1})$ meson also couples to a parity partner 
state, which is in this case the $({\bf 1}\otimes\gamma_4\gamma_5)$
scalar meson. The parity partner of the 
$(\gamma_5\otimes{\bf 1})$ however is exotic and therefore makes no
contribution to the $(\gamma_5\otimes{\bf 1})$ propagator measured on the 
lattice. 
Furthermore, a variance reduction trick we discuss in Section \ref{VR} applies 
only to the $(\gamma_5\otimes{\bf 1})$ operator. For these reasons we use 
the $(\gamma_5\otimes{\bf 1})$ state exclusively in this work.

To apply the $\Delta_{\gamma_5\otimes{\bf 1}}$ operator to a source vector or 
fermion propagator, we covariantly and symmetrically shift the source by one 
lattice unit in each of the four dimensions. We then apply the appropriate 
Kogut-Susskind phases \cite{Golterman:1985dz}, the exact formulation of this 
phase factor depending on the gamma-matrix conventions used in the 
simulation code.

Disconnected correlators are by nature noisy; they are directly sensitive 
to the fluctuations in the fermionic and gluonic sea. Consequently it is 
essential to extract more measurements from each configuration than 
can be gleaned from a 
single point source inversion. 
In order to achieve this we use the stochastic source method \cite{Bernardson:1993yg, Dong:1993pk, Wilcox:1999ab, Farchioni:2004ej}. 
A set of $N_{\rm src}$ independent noise source vectors $\{\eta^a\}$ 
($a=1,2,\dots N_{\rm src}$)
is chosen and normalised
so that it has the orthogonality property
\begin{equation}
\langle\eta^{a\dagger}_i\eta^b_j\rangle_\eta =\delta_{ab}\delta_{ij}
\label{etaorthog}
\end{equation}
where $\langle\rangle_\eta$ denotes the expectation value over all random noise $\eta$
and $i$ labels generically the appropriate vector components such as space, Euclidean time, colour etc.
Averaging over a sufficiently large number of 
samples $N_{\rm src}$, the set of noise vectors then approximates the desired expectation value (\ref{etaorthog}) and so
reliable unbiased estimators for operators sums can be constructed. 
On a given configuration, and on each time slice $t$
we calculate
\begin{equation}
\label{loop_op}
{\mathcal O}_{\gamma_5\otimes{\bf 1}}(t)=\sum_{i\in t}\langle\eta_i^{\dagger}
 \Delta_{\gamma_5\otimes{\bf 1}}M_{ij}^{-1}\eta_j\rangle_\eta,
\end{equation}
where the sum over $i$ is restricted to the subset of lattice points on 
timeslice $t$ and, from now on, $\langle\rangle_\eta$ denotes the expectation value 
estimated from the average over $N_{\rm src}$ noise vectors.

The Kentucky group has found analytically that among real noise sources,
$Z(2)$ noise sources should 
offer the minimal variances for determining propagators~\cite{Dong:1993pk}. 
We initially tested both complex $Z(2)$ and Gaussian noise sources on a small 
number of $16^3\times 32$  $N_f=2$ lattices and found 
that Gaussian noise sources produced slightly smaller errors than
$Z(2)$ noise sources (see \cite{Gregory:2005me}). Based on 
this preliminary result we employed Gaussian volume sources for further 
simulation.

\begin{figure}[t]
\resizebox{5.0in}{!}{\includegraphics{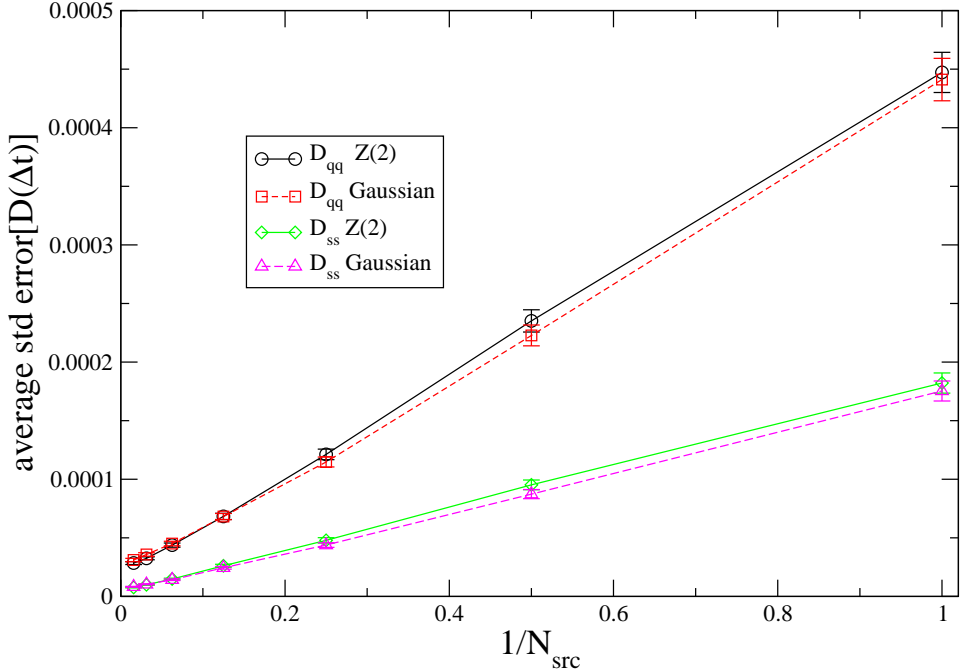}}
\caption{A comparison of Gaussian and $Z(2)$ noise sources for 67 $20^3\times 64$ $N_f=2+1$
lattices for $\beta=6.76$ $am=0.01,0.05$. The 
plot shows the 
standard error on the  disconnected correlator $D(\Delta t)$, averaged over 
$\Delta t$, versus inverse number of noise sources for the standard 
$(\gamma_5\otimes{\bf 1})$ operator.
\label{Z2gaussfig_g5x1}}
\end{figure}

To investigate somewhat more thoroughly the apparent discrepancy between 
this and the Kentucky group's result we looked at a larger sample ($N_\mathrm{cfg}=67$) of
$2+1$-flavour $\beta=6.76$ lattice configurations, and measured the light quark
disconnected correlator $D_{qq}(\Delta t)$ using both $Z(2)$ and 
Gaussian noise source 
vectors. Since the size of the errors of disconnected correlators is found to be almost
independent of the magnitude of the correlator itself, we averaged the
error over the timeslice separation $\Delta t$ allowing us to quote 
``errors on the errors''. In Fig. \ref{Z2gaussfig_g5x1} the top two curves
are the errors on the $\gamma_5\otimes{\bf 1}$ disconnected correlator
measured with Gaussian noise (solid line) and with $Z(2)$ noise (dashed
line) as a function of $1/N_{\rm src}$. It is not apparent that either type of noise
presents a large advantage over the other in the relevant region of 
$N_{\rm src}\geq 8$.

These figures display the interesting property, previously noted in
\cite{Michael:1998sg}, 
that the errors of disconnected correlators measured with the volume noise 
source method appear to be inversely proportional to $N_{\rm src}$, rather 
than to $\sqrt{N_{\rm src}}$ as one might naively assume. In \cite{Foley:2005ac}
Foley et al. attribute this to a property of the dilution method, however we 
find it to be a generic property of volume sources as well. This can best 
be understood by considering that while the loop operator  on a given 
configuration  (\ref{loop_op})
\begin{equation}
{\mathcal O}_{\gamma_5\otimes{\bf 1}}(t)=
\frac{1}{N_{\rm src}}\sum_{a=1}^{N_{\rm src}}\left[\sum_{i,j\in t}\eta_i^{a\dagger}
 \Delta_{\gamma_5\otimes{\bf 1}}M_{ij}^{-1}\eta_j^a\right],
\end{equation}
has $N_{\rm src}$ terms in the sum over noise sources, the disconnected correlator
on that configuration has $N_{\rm src}^2$ terms:
\begin{eqnarray}
{\mathcal D}_{\gamma_5\otimes{\bf 1}}(\Delta t)&=&
{\mathcal O}_{\gamma_5\otimes{\bf 1}}(t){\mathcal O}_{\gamma_5\otimes{\bf 1}}(t+\Delta t)\nonumber\\
&=&
\frac{1}{N_{\rm src}^2}\sum_{a=1}^{N_{\rm src}}\sum_{b=1}^{N_{\rm src}}\left[\sum_{i,j\in t}\eta_i^{a\dagger}
 \Delta_{\gamma_5\otimes{\bf 1}}M_{ij}^{-1}\eta_j^a\right]
\left[\sum_{k,l\in t+\Delta t}\eta_k^{b\dagger}
 \Delta_{\gamma_5\otimes{\bf 1}}M_{kl}^{-1}\eta_l^b\right],
\end{eqnarray}
so the variance likewise decreases like $1/N_{\rm src}^2$.

\begin{figure}[tbh]
\resizebox{5.0in}{!}{\includegraphics{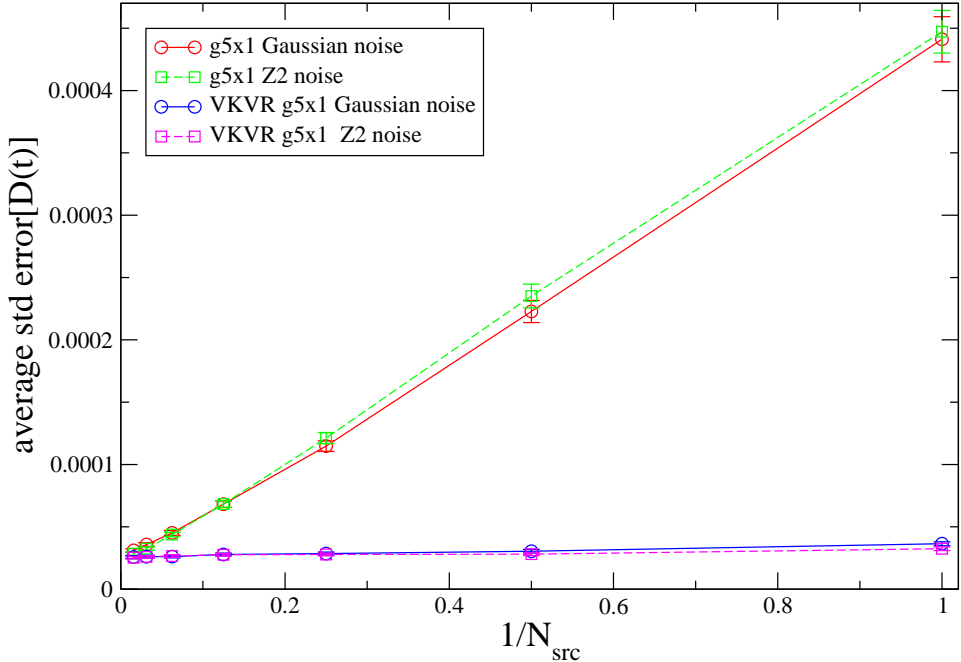}}
\caption{The dependence of the error on $D_{qq}(t)$ with $1/N_{\rm src}$ for the 67
lattice configurations described in Fig.~\ref{Z2gaussfig_g5x1}. 
The comparison is for $Z(2)$ and Gaussian
noise for both standard and VKVR-improved $(\gamma_5\otimes{\bf 1})$ operators (see text).
\label{gauss_Ns_dependence_g5x1_g5x1kcp_compare}}
\end{figure}

\begin{figure}[tbh]
\resizebox{5.0in}{!}{\includegraphics{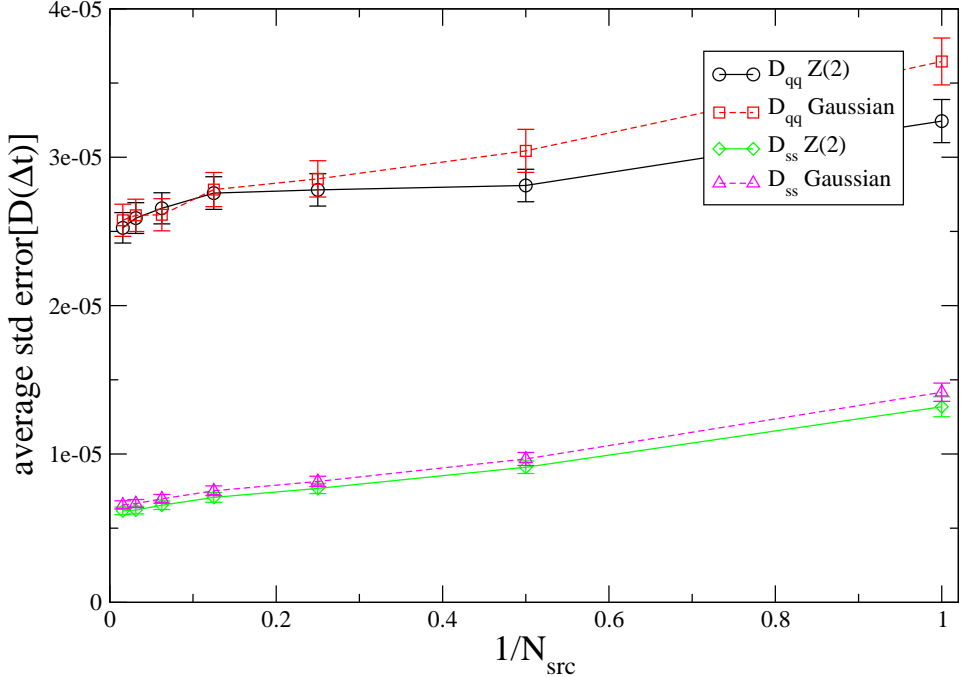}}
\caption{Dependence of the averaged standard error of $D_{qq}(t)$ and 
$D_{ss}(t)$ on $1/N_{\rm src}$ for the $67$ lattice configurations described in Fig..~\ref{Z2gaussfig_g5x1}. 
VKVR-improved $(\gamma_5\otimes {\bf 1})$ operators are used.
\label{gauss_Ns_dependence_kcp}}
\end{figure}

\begin{figure}[t]
\resizebox{5.0in}{!}{\includegraphics{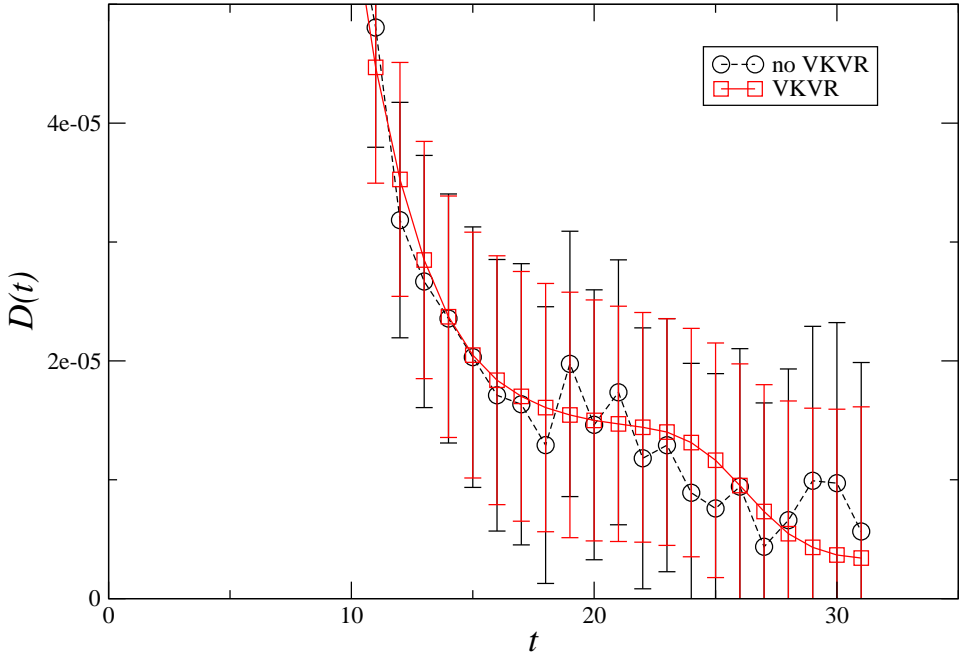}}
\caption{A comparison of disconnected correlators for $\beta=6.76$ $am=0.01$
obtained with and without using the 
VKVR variance reduced operators for $N_{\rm src}=64$.
\label{VKVRfig}}
\end{figure}

\begin{figure}[tbh]
\resizebox{5.0in}{!}{\includegraphics{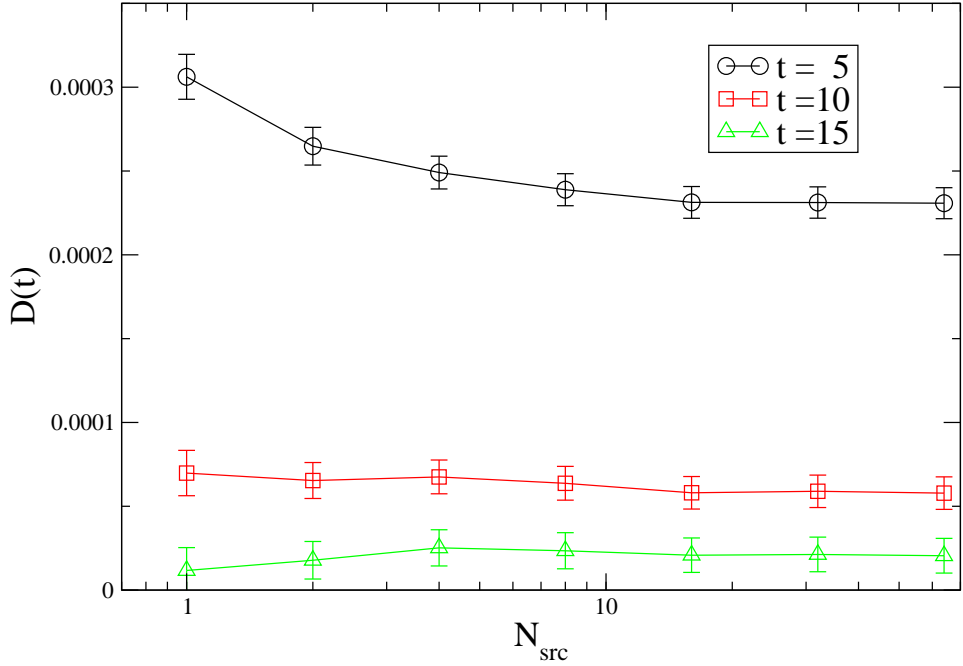}}
\caption{Dependence of $D_{qq}(t)$ on $N_{\rm src}$ for 658
$\beta=6.76$ $am=0.01$ $20^3\times 64$ lattices.
\label{Dqq_Ns_dependence}}
\end{figure}

\begin{figure}[tbh]
\resizebox{5.0in}{!}{\includegraphics{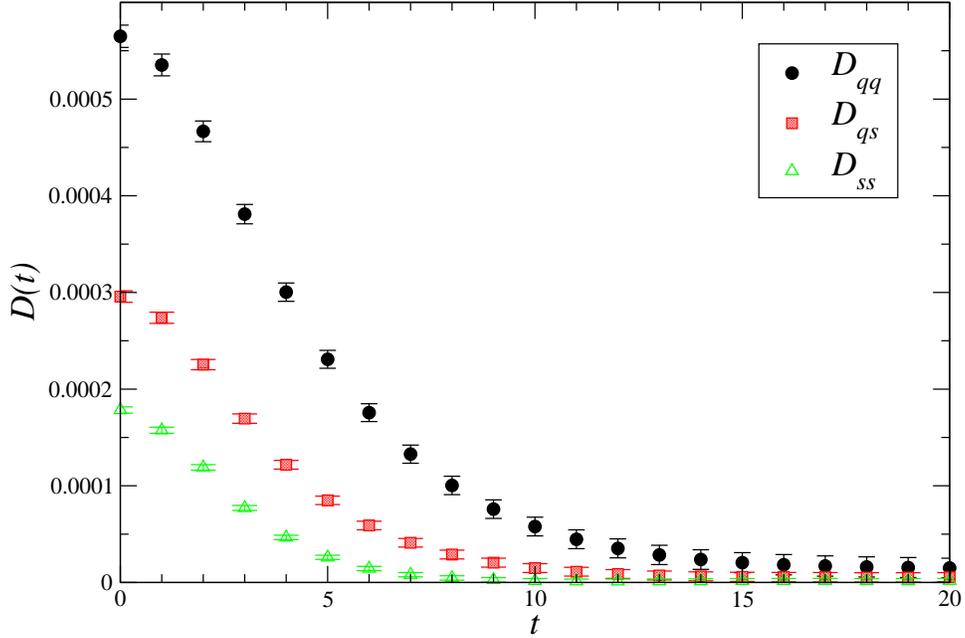}}
\caption{Disconnected correlators for 658
$\beta=6.76$ $am=0.01$ $20^3\times 64$ lattices.
\label{all_Dcorrs_b676m010m050}}
\end{figure}

\subsection{Variance reduction}\label{VR}

Results from the TrinLat collaboration, using Wilson like fermions,
 indicate that variance is reduced by 
using ``diluted'' noise sources when using stochastic sources to measure 
connected correlators~\cite{Foley:2005ac}. 
As the name implies, such diluted sources are non-zero 
only on some subset of the
lattice. It is then important to use a set of sources for which the non-zero
subsets together span the entire lattice volume. The subsets may, for example, 
be timeslices, colors, or hypercube corners. We tested several noise dilution
schemes 
in the present application and found no advantage over using 
traditional volume-filling sources (see \cite{Gregory:2005vy}). 

In \cite{Venkataraman:1997xi}, Venkataraman and Kilcup introduce a variance 
reduction technique that is applicable for the the $(\gamma_5\otimes{\bf 1})$
staggered meson operator. The Venkataraman-Kilcup variance reduction
(VKVR) trick uses the fact that the staggered Dirac operator $M=(\Dslash + m)$
and its Asqtad-improved variants (and their inverses) connect sites 
separated by an odd number of links. Likewise $M^\dagger M$ and 
its inverse connect sites separated by an even number of gauge links. Here
we recall that the quark and anti-quark are displaced from each other 
by four links by the $(\gamma_5\otimes{\bf 1})$ operator. Now
\begin{equation}
\eta_x M^{-1}\eta_y \nonumber
\end{equation}
is clearly
\begin{equation}
\eta_x (M^\dagger M)^{-1}M^\dagger\eta_y = \eta_x  (M^\dagger M)^{-1} (-\Dslash +m)
\eta_y\, .
\end{equation}

However, since the sites denoted by $x$ and $y$ (quark and antiquark) are 4 links apart, the term
proportional to $\Dslash$ has zero expectation value so
\begin{equation}
\langle\eta_x M^{-1}\eta_y \rangle_\eta= m\langle\eta_x(M^\dagger M)^{-1}\eta_y\rangle_\eta\, .
\label{VKVR_trick}
\end{equation}
It turns out, however, that the right-hand expression has a significantly decreased variance
with respect to stochastic noise.
There is no additional computational cost since we are already computing 
$\phi = M^{-1}\eta$, and we can express the right hand side of 
(\ref{VKVR_trick}) as $m\langle\phi^\dagger\phi\rangle$.

As the VKVR trick should be applicable to disconnected loops for
any meson operator with the quark and antiquark separated by an
even number of lattice spacings, for example the $({\bf 1}\otimes{\bf 1})$
scalar meson.

The error on elements of the disconnected correlators is composed of a 
stochastic part --- which decreases like $1/N_{\rm src}$ as mentioned above --- and
a gauge part, which remains after $N_{\rm src} \rightarrow \infty$ and decreases 
as $1/\sqrt{N_{\rm cfg}}$.
The effect of the VKVR trick on disconnected correlator errors is dramatic.
Fig.~\ref{gauss_Ns_dependence_g5x1_g5x1kcp_compare} shows the error on the 
disconnected correlator (averaged over time separations) as a function of the 
inverse number of noise sources. With just a few noise sources the VKVR
trick nearly eliminates all of the stochastic component of the error on $D_{qq}$.
At $N_{\rm src}=64$ the magnitude of the errors on the VKVR and normal 
$\gamma_5\otimes{\bf 1}$ disconnected correlators become comparable 
(Fig. \ref{gauss_Ns_dependence_g5x1_g5x1kcp_compare} and Fig. \ref{VKVRfig}).

To calculate the connected and disconnected correlators of the 
$\gamma_5\otimes{\bf 1}$ state we wrote specialised routines for the 
Chroma software system \cite{Edwards:2004sx}.

On each configuration, for the light and strange quark masses, we calculated 
the connected correlator for the 
singlet pion $(\gamma_5\otimes{\bf 1})$. We also calculated the mean of 
${\rm Tr}\left[\Delta_{\gamma_5\otimes{\bf 1}}M^{-1}\right]$ on each timeslice using $N_{\rm src}=64$
Gaussian noise sources for each. The latter we use to calculate disconnected 
correlators. We then use all of these quantities to make the $D/C$ ratio
as in Equation \ref{nondegenerate_ratio_eq}. 
We fit $R(t)$ to the functional form in (\ref{degenerate_ratio_eq}) to
determine the splitting between the pion and singlet masses.

\section{Results} \label{se:results}

\subsection{Quenched analysis}

With the availability of improved algorithms and significantly more
powerful computational resources, 
the quenched approximation to lattice QCD no longer plays
such a significant role.
However, it remains instructive to study
singlet pseudoscalar mesons from quenched lattice QCD calculations.
In particular the Witten-Veneziano relation
relates the topological susceptibility, in the large $N$
limit of the $SU(N)$ pure gauge theory, to a contribution
to the mass of the $\eta^\prime$ meson.
A comparison between quenched and full-QCD systems highlights the effect of 
dynamical sea quarks. This is particularly instructive in 
studying flavour singlet mesons. Further, since quenched configurations are 
cheap to generate, one can use very large ensembles to illuminate the
nature of the statistical fluctuations of singlet quantities and 
possible long auto-correlation times associated with
slow topological modes.

\begin{figure}[tbh]
\resizebox{6.4in}{!}{\includegraphics{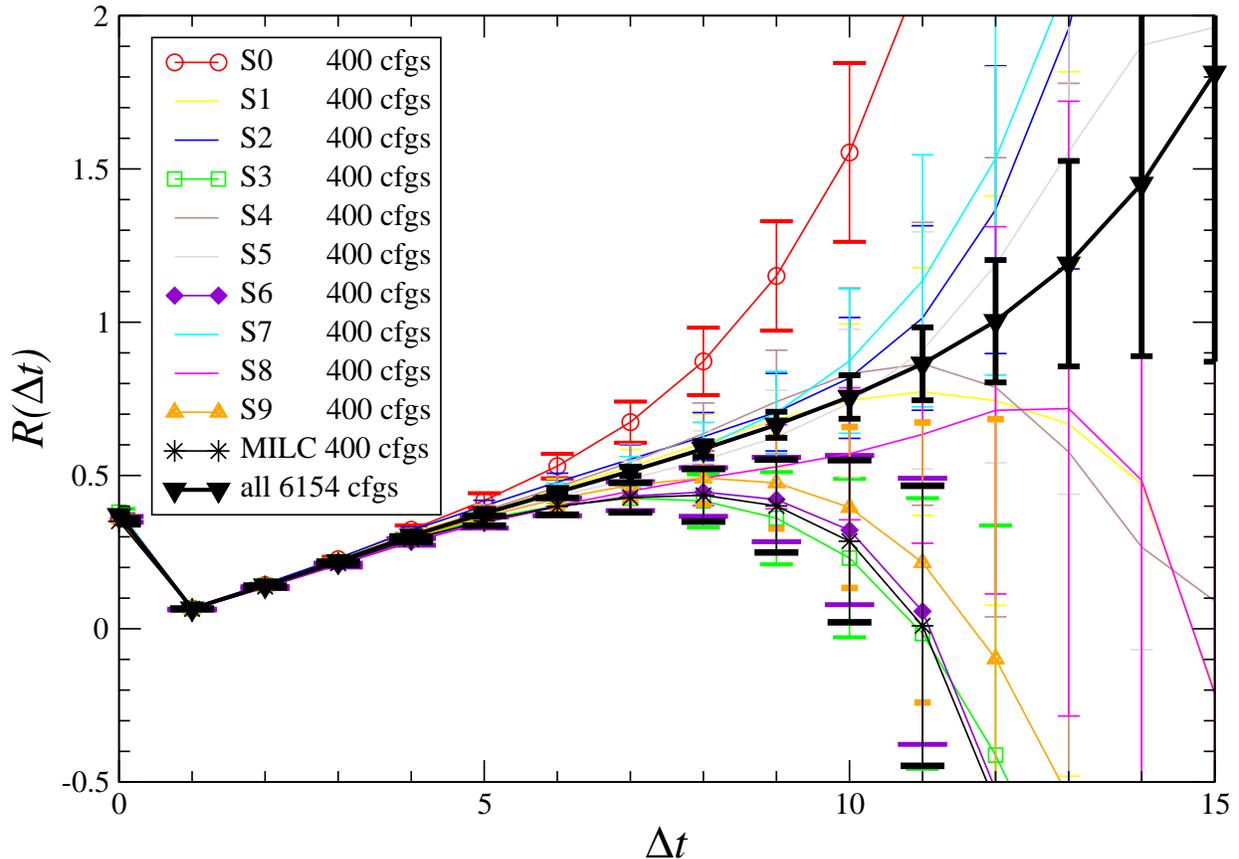}}
\caption{D/C ratio for 6154 $\beta=8.00$ quenched lattices with valence quark 
mass $am=0.05$, and for 11 subsets of 400 configurations. We highlight five of 
these subsets, including the original MILC configurations, as disagreeing with 
the mean by more than one $\sigma$.  \label{binned_dc_rat_b800}}
\end{figure}

We calculated connected and 
disconnected correlators on the quenched gauge configurations
at $\beta=8.00$ (neglecting 8 thermalisation configurations)
with a valence mass of 0.05. In 
Figure~\ref{binned_dc_rat_b800} the ratio $R(t)$
is plotted for all the quenched configurations, as well
as for subsets of the quenched configurations.
Quenched staggered
QCD involves no determinant fourth-roots so there are no obvious theoretical 
reasons for a significant deviation from (\ref{quenched_ratio_eq})
for the MILC ensemble, when
lattice artifacts and finite size effects are neglected. The original MILC coarse 
quenched ensemble ($\beta=8.00$) had 408 configurations, and upon
extending the quenched ensemble as described in Section \ref{sim_meas} and
Table \ref{extended_quenched_configs}, we see from
Fig.~\ref{binned_dc_rat_b800} that the $D/C$ ratio
calculated with the full 6154 configurations is more linear and, as such, in better
agreement with the expression (\ref{quenched_ratio_eq}) for
$\Delta t\leq 15$.

\begin{figure}
\begin{center}
\resizebox{5.0in}{!}{\includegraphics{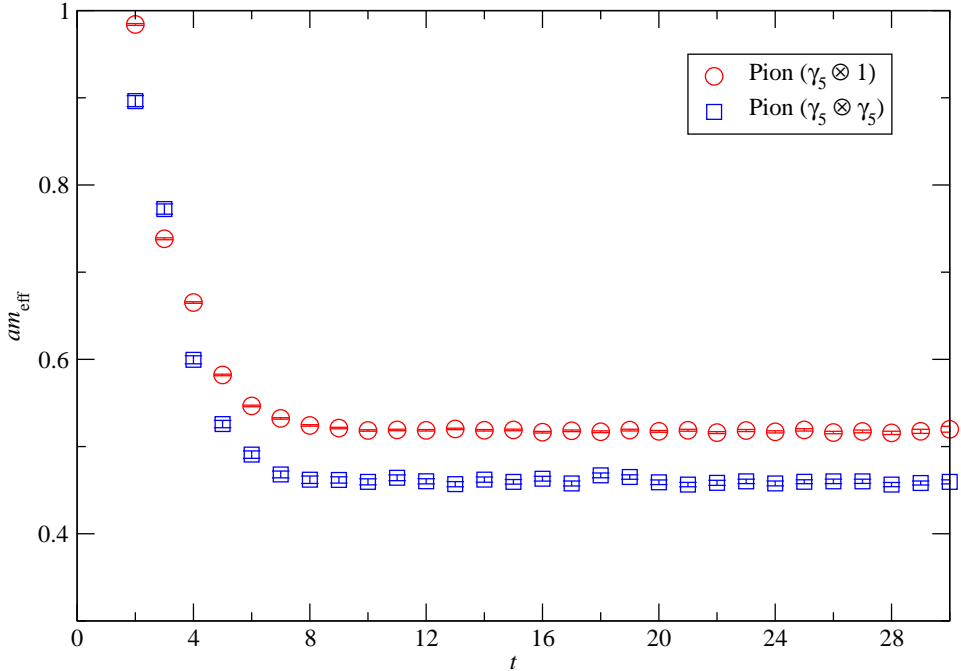}}
\end{center}
\caption {
Effective masses for the $\gamma_5 \otimes \gamma_5 $
and $\gamma_5 \otimes {\bf 1} $ pseudoscalar mesons
for the quenched data set ($\beta=8.0$).
}
\label{fig:meffQUENCHED}
\end{figure}

In Fig.~\ref{fig:meffQUENCHED} we show the effective
mass plots for the light $\gamma_5 \otimes {\bf 1} $
and $\gamma_5 \otimes \gamma_5 $ non-singlet pseudoscalar mesons.
We computed correlators for the
$\gamma_5 \otimes \gamma_5 $ pseudoscalar mesons only
on the original data set, but computed 
the $\gamma_5 \otimes {\bf 1} $ connected and disconnected correlators 
on the full extended data set.

In Table~\ref{tab:massRESULTS} we report the masses 
for the $\gamma_5 \otimes {\bf 1}$ and $\gamma_5 \otimes \gamma_5$
pseudoscalar mesons. The fits included the full correlation
matrix in time and we report both one- and two-cosh\ mass fits.
The MILC collaboration obtained the mass $0.46043(20)$ for
the mass of the $\gamma_5 \otimes \gamma_5 $ pseudoscalar
meson, using Coulomb gauge fixed wall sources \cite{Bernard:2001av}. This is
in good agreement with our value in Table~\ref{tab:massRESULTS}.
As expected, the 
inclusion of the second state allows us to fit 
much closer to the origin~\cite{Aubin:2004wf}.

\begin{table}[tb]
  \caption{
Masses for the 
light pseudoscalars mesons 
(using connected correlators only) 
from the 
quenched data set ($\beta=8.00$).
The $\gamma_5 \otimes {\bf 1}$ correlators were
measured with 8 times the statistics of the $\gamma_5 \otimes \gamma_5$ channel.
}
\begin{center}
\begin{tabular}{|c|c|c|c|c|} \hline
Channel & t-region & $a m_1$ & $a m_2$  & $\chi^2/dof$ \\ \hline
$\gamma_5 \otimes \gamma_5 $ & 13-26 & 0.4601(6) & - & 16.8/ 12  \\ 
$\gamma_5 \otimes \gamma_5 $ & 5-25 & 0.4603(6) & 1.88(56) & 23.1/17  \\ 
$\gamma_5 \otimes {\bf 1}$ & 11-24 & 0.5182(3) & - & 14.2 / 12  \\ 
$\gamma_5 \otimes {\bf 1}$ & 5-23 & 0.5180(3) & 1.27(5) & 17.1/16  \\ 
\hline
\end{tabular}
\end{center}
\label{tab:massRESULTS}
\end{table}

Quenched chiral perturbation theory~\cite{Venkataraman:1997xi,Bernard:1992mk}
makes predictions for the ratio of 
disconnected and connected correlators:
\begin{equation}
R(t) = \frac{(m_0^2 - \alpha m_{NP}^2) } {2 m_{NP}}t 
+ \frac{m_0^2 + \alpha m_{NP}^2}{ 2 m_{NP}^2} 
\label{RquenchPerturb}
\end{equation}
where $\alpha$ is the parameter of the kinetic
term of singlet pseudoscalar meson~\cite{Bernard:1992mk},
and $m_0$ is the difference between the masses $m_{SP}$
and  $m_{NP}$. Comparing (\ref{RquenchPerturb}) with
 (\ref{quenched_ratio_eq}) we see that it defines the 
parameters $A^\prime$ and $B^\prime$ in (\ref{quenched_ratio_eq}).

The value of $\alpha$ is expected to be small, so
we neglect it~\cite{Venkataraman:1997xi,Bernard:1992mk} --- indeed
Bardeen et al.~\cite{Bardeen:2003qz}
estimate that $\alpha$ = $0.03 \pm 0.03$
and advocate setting $\alpha$ = 0.

Fitting the quenched $D(t)/C(t)$ data to the model in 
(\ref{quenched_ratio_eq}),
we obtain 
a slope of $B^\prime=0.068(3)$. Using the mass of the $\gamma_5 \otimes {\bf 1}$
pseudoscalar meson from Table~\ref{tab:massRESULTS}
and $B^\prime=\frac{m_0^2}{ 2 m_{NP}^2} $,
we obtain $m_0 = 0.76(2)$ GeV, where we have multiplied
by $\sqrt{n_f}$ = $\sqrt{3}$~\cite{Kuramashi:1994aj}.

\begin{figure}
\begin{center}
\resizebox{5.0in}{!}{\includegraphics{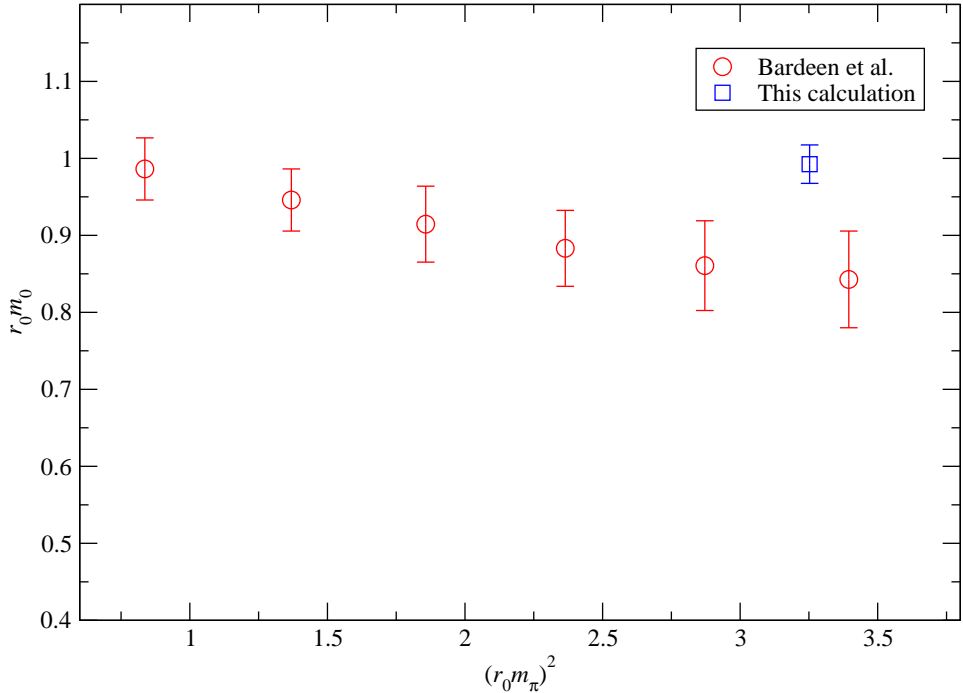}}
\end{center}
\caption {
Mass dependence of the $m_0$
from Bardeen et al.~\cite{Bardeen:2003qz}
and this calculation in units of $r_0$.
There are no factors of $n_f$ included.
}
\label{fig:m0MASSdepend}
\end{figure}

In Fig.~\ref{fig:m0MASSdepend} our value for $m_0$
and the values from
Bardeen et al.~\cite{Bardeen:2003qz,Bardeen:2004md}
are plotted
against the square of the pion mass in units of $r_0$.
The results from Bardeen et al.~\cite{Bardeen:2003qz,Bardeen:2004md}
at $\beta$ = 5.9 were used with the  value of $r_0/a$ from
the ALPHA collaboration~\cite{Guagnelli:1998ud}.


Estimates for the 
value of $m_0$, in the chiral limit, can also be obtained from the 
Witten-Veneziano relation
\begin{equation}
m_0^2 = \frac{4 N_f \chi_T }{f_\pi^2}
\label{eq:WV}
\end{equation}
where the topological susceptibility $\chi_T$
is evaluated
in the large $N_c$ limit of the pure gauge theory
and using $f_\pi$ with a  normalisation of 132 MeV.
Shore~\cite{Shore:2006mm} describes a modern approach to the use of 
the topological susceptibility in the study of 
$\eta$ and $\eta'$ mesons.
Lucini and Teper ~\cite{Lucini:2001ej} have shown
that the topological susceptibility is only weakly dependent
on the number of colours.

The MILC collaboration~\cite{Bernard:2003gq} 
obtained $\chi_T r_0^4 = 0.0543(28)$ at $\beta = 8.00$
, and $\chi_T r_0^4 = 0.0569(26)$  at $\beta = 8.40$. 
These numbers are in good agreement with other
quenched results such as those by Del Debbio et al.~\cite{DelDebbio:2004ns},
who obtain $\chi_T r_0^4 = 0.059(3)$ in the continuum limit.
There has also been a recent calculation by Durr et al.~\cite{Durr:2006ky},
who obtain
$\chi_T r_0^4 = 0.0524(7)(6)$  in the continuum limit.

Using $\chi_T r_0^4 = 0.0543(28)$ in (\ref{eq:WV})
with $N_f = 3$, and taking the value of $r_0 = 0.467$ fm
from the HPQCD and MILC collaborations, one obtains
$m_0^2 = ( 1.09\, {\rm GeV} )^2$. Using the value of 
$r_0 = 0.5$ fm, instead, one would obtain 
$m_0^2 = ( .952\, {\rm GeV} )^2$.

The value of $m_0$ can also be estimated 
from~\cite{Donoghue:1992dd}, assuming
that the $\eta$ is a pure SU(3) octet and the 
$\eta^\prime$ is the SU(3) singlet.
\begin{equation}
m_0^2 = m_{\eta'}^2 + m_{\eta}^2 - 2 m_K^2
\end{equation}
This method gives a value of $m_0^2$ = $(0.854\, {\rm MeV} )^2$.


The mass dependence of the $m_0$ parameter 
has recently been discussed by 
Sharpe~\cite{Sharpe:2003hw}.
In quenched lattice calculations,  Kuramashi et al.~\cite{Kuramashi:1994aj}
found that the $m_0$ parameter had a negative slope with 
respect to the quark mass where it is more usual to see a 
positive slope (see the results by MILC~\cite{Aubin:2004wf}).
Sharpe's predictions were tested by Bardeen
et al.~\cite{Bardeen:2004md}. Liu and Lagae~\cite{Lagae:1994bv}
found that a novel mass-dependent renormalisation factor
removed the bulk of the mass dependence of $m_0$ observed in
the study by Kuramashi et al.~\cite{Kuramashi:1994aj}.

Our result $m_0 = 0.76(2)$ GeV described above, 
with the valence quark mass set approximately to 
that of the strange quark,
is consistent with other estimates. A chiral extrapolation
would be required to make a more precise comparison.

\subsection{Unquenched analysis} \label{unquence_sec}

\begin{figure}
\begin{center}
\resizebox{5.0in}{!}{\includegraphics{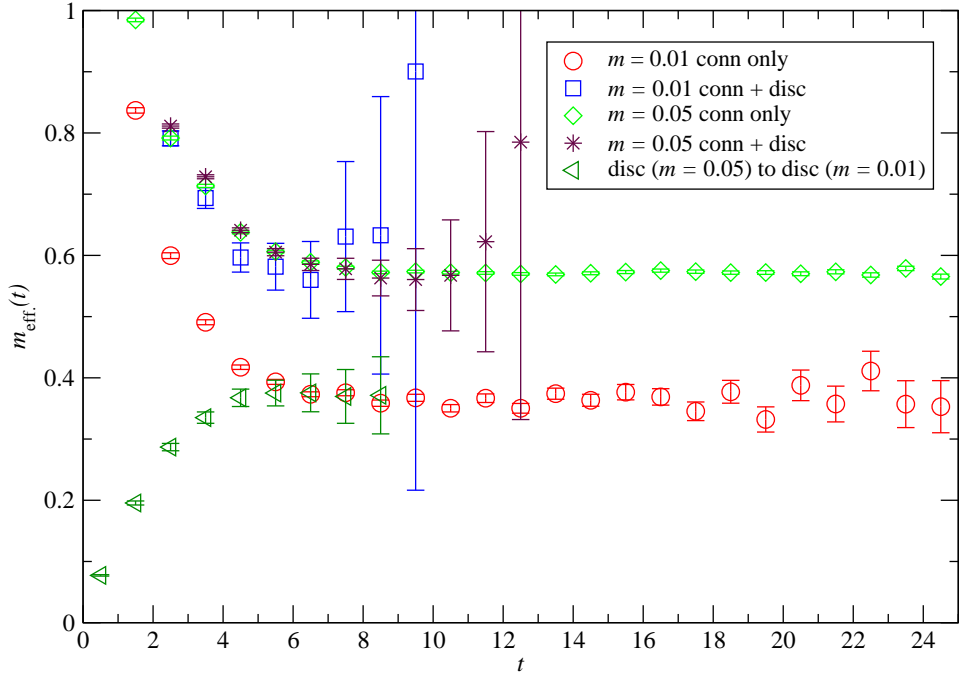}}
\end{center}
\caption {
Effective masses for the $(\gamma_5 \otimes {\bf 1})$ channel
for the light and strange quarks 
for the $\beta$ = 6.76, m=0.01/0.05 data set.
}
\label{fig:meffB676}
\end{figure}

In Fig.~\ref{fig:meffB676} we present effective mass plots for the
$(\gamma_5 \otimes {\bf 1})$ correlators, with the connected and sum of
connected and disconnected correlators, for the $\beta$=6.76,
$m_q/m_s$ = 0.01/0.05 data set. The effective mass from
a correlator of a disconnected loop at the light quark mass with a
disconnected loop at the strange quark mass
is also included in Fig.~\ref{fig:meffB676}.

In Table~\ref{tab:massRESULTSB676CON} we report single- and 
double-exponential fits to the connected $\gamma_5 \otimes {\bf 1}$ pseudoscalar
correlators. MILC obtained the masses of the
 $\gamma_5 \otimes \gamma_5$ pseudoscalar mesons
to be 0.22446(22) and 0.49443(25) in lattice units for the 
light and strange quarks respectively~\cite{Bernard:2001av}.
At this lattice spacing the light $\gamma_5 \otimes {\bf 1}$ pseudoscalar meson
is split from the $\gamma_5 \otimes \gamma_5 $ pseudoscalar meson 
by about 200 MeV. This mass splitting is  caused  by
taste violations and is expected to go to zero as the
continuum limit is taken.
Indeed, the MILC collaboration  
has presented evidence that the 
mass splittings between different pseudoscalar mesons
go like
$O(\alpha_s^2 a^2)$, as expected, for this action~\cite{Aubin:2004wf}.

If the mass splitting between the 
connected $\gamma_5 \otimes {\bf 1}$ pseudoscalar mesons
and the $\gamma_5 \otimes \gamma_5$ correlator was
taken as a systematic error at this lattice spacing, then
this would imply the error in the $\eta$ mass at this 
lattice spacing is roughly 30\%. 

In the twisted mass formalism there is a flavour symmetry breaking
term that causes a mass splitting between the $\pi^0$ and $\pi^+$
mesons at non-zero lattice spacing.  
Although the physics of flavour
symmetry breaking is different for 
staggered fermions~\cite{Lepage:1998vj}
and the twisted
formalism, it is encouraging that experience with the twisted mass
formalism has shown that the disconnected diagrams reduce the mass
splitting between the masses of the $\pi^0$ and $\pi^+$ mesons, caused
by isospin violation outside the continuum limit, over the estimate
from the connected correlators~\cite{Jansen:2005cg}.

We see from Table~\ref{tab:massRESULTSB676DIS} that consistent ground
states are determined from the $\overline{s}\gamma_5 s$ and 
$\overline{q}\gamma_5 q$
operators when the disconnected diagrams are included.
This is non-trivial, particularly because the masses from
the connected correlators
are very different for light and strange quarks. 
This is true because the physical $\eta$ contains
both $\overline{s}\gamma_5 s$ and $\overline{q} \gamma_5 q$ 
quark content, so either
interpolating operator should couple to the $\eta$ meson
and produce the same ground state mass.
The 
$\eta - \eta^\prime$ mixing is further
discussed in Section \ref{se:theory}.
The recent CP-PACS/JLQCD calculations enforced a consistent mass
for the $\overline{s}s$ and $\overline{q}q$ operators using their
variational method~\cite{Aoki:2006xk}. The contribution from the 
disconnected part of the correlators appears to compensate for the 
increase of the masses from the connected part.

The correlator between a light and strange pseudoscalar
loop should also couple to the $\eta$ and $\eta^\prime$ mesons.
Fig.~\ref{fig:meffB676} shows that the correlator between a
light quark loop and strange quark loop 
(off-diagonal part of (\ref{matrix_prop})) 
seems to have a plateau
at a mass different to the value from 
light or strange correlators (diagonal parts of (\ref{matrix_prop})).
It is not clear why this happening --- Aubin and Bernard
have published propagators for the neutral $\eta$ from
staggered chiral perturbation theory~\cite{Aubin:2003mg},
but unfortunately these do not include terms for $\eta-\eta'$ mixing
that are essential for this analysis.
One obvious possibility is that the ground state
is not being isolated for the singlet correlators in
Fig.~\ref{fig:meffB676}. It has been noted before
by the SESAM collaboration and 
Bardeen et al.~\cite{Bardeen:2003qz,Bardeen:2004md}
that the majority of the excited state contamination in
a flavour singlet pseudoscalar correlator is from the connected 
part of the correlator. Hence the correlator of a strange loop
with a light quark loop could approach the ground state 
at a different rate to correlators with a connected
contribution. Given that the signal for correlators that
include a disconnected contribution in Fig. \ref{fig:meffB676}
dies in the noise beyond timeslice 7, it is difficult to
tell where the true plateau lies.

We tried the variational analysis discussed in
Section~\ref{se:fittvary}. The $\chi^2/dof$ of the
fits were larger than 7. The problem is, as we discussed
above, that the correlators for a light loop to strange loop
seem to be plateauing at different mass to the
singlet light and strange loops.
The variational fitting formula (\ref{eq:varyFIT}) assumes
that each element of the smearing matrix couples to the
same ground state. We also tried looking at eigenvalues
of the matrix, but the two eigenvalues just reproduced
the diagonal and off diagonal correlators. One possibility
would be that there is a normalisation problem between
the disconnected and connected loops. However the magnitude of the
light disconnected loop is much larger than that of the strange
disconnected loop, so a consistent change in the normalisation
of the disconnected loops makes it hard to get the effective mass of 
the light-light
correlator to plateau at the same mass as the strange-strange
correlator. This is also a prerequisite for the variational
analysis.

The analysis of the non-singlet $a_0$ correlator using improved
staggered fermions turned out to be possible, but
non-trivial~\cite{Prelovsek:2005rf,Bernard:2006gj}.  In this case the
ground state for the singlet pseudoscalar channel is the $\eta$ that
is stable under the strong interactions.

\begin{table}[tb]
  \caption{
Masses for the light pseudoscalars mesons for the unquenched
data sets obtained from
($\gamma_5 \otimes {\bf 1}$)
connected correlators.
}
\begin{center}
\begin{tabular}{|c|c|c|c|c|c|c|c|} \hline
$\beta$ & $m_l/m_s$ & $m_v$  & region & $a m_1$ & $a m_2$  & $\chi^2/dof$ \\
\hline
6.76 & 0.01/0.05    &0.05  & 10-17 & 0.572(1) & - & 5.5/6  \\
6.76 & 0.01/0.05    &0.05   & 5-11 & 0.571(2) & 1.27(11) & 3.5/3  \\
6.76 & 0.007/0.05   & 0.05 & 11-16 & 0.566(2) & - & 2.6/4  \\
6.76 & 0.007/0.05   & 0.05 & 5-11 & 0.567(3) & 1.19(13) & 4.6/3  \\
6.76 & 0.01/0.05    &0.01   & 11-16 & 0.363(4) & - & 4.8/4  \\
6.76 & 0.01/0.05    &0.01   & 4-11  & 0.358(5) & 1.00(12) & 4.3/4  \\
6.76 & 0.007/0.05   &0.007  & 11-16 & 0.331(6) & - & 0.6/4  \\
6.76 & 0.007/0.05   &0.007  & 5-11 & 0.33(1) & 0.91(34) & 2.9/3  \\
6.85 & 0.05/0.05 &0.05 & 12-17 & 0.553(2) & - & 3.8/4  \\
6.85 & 0.05/0.05 &0.05 & 4-14 & 0.554(2) & 1.33(8) & 8.7/7  \\
\hline
\end{tabular}
\end{center}
\label{tab:massRESULTSB676CON}
\end{table}

\begin{table}[tb]
  \caption{
Masses for the light ($\gamma_5 \otimes {\bf 1}$) 
pseudoscalars mesons for the unquenched
data sets, from the correlators that includes the disconnected 
diagram.
}
\begin{center}
\begin{tabular}{|c|c|c|c|c|c|} \hline
$\beta$ & $m_q/m_s$ & $m_v$  & region & $a m_1$  &  $\chi^2/dof$ \\ \hline
6.85 & 0.05/0.05  & 0.05 & 7-11 & 0.64(16) &  1.6/3  \\ 
6.76 & 0.01/0.05  & 0.05 & 5-11 & 0.59(2) &  0.4/3  \\ 
6.76 & 0.007/0.05 & 0.05 & 5-9  & 0.59(2) &  1.4/4  \\ 
6.76 & 0.01/0.05  &0.01  & 4-9 & 0.58(6) &  0.6/3  \\ 
6.76 & 0.007/0.05 &0.007 & 5-9 & 0.64(9) & 1.4/3   \\ 
\hline
\end{tabular}
\end{center}
\label{tab:massRESULTSB676DIS}
\end{table}

The numbers for the lightest mass in 
Table~\ref{tab:massRESULTSB676DIS}
are about 0.6 in lattice units with weak quark mass
dependence. This corresponds
to a mass in physical units of 990 MeV ---
the ground state in this channel should be 
the $\eta$ with a mass of 548 MeV. The masses
in Table~\ref{tab:massRESULTSB676DIS} are probably
contaminated by excited state contributions, so 
are only an upper limit. A fit to the  
mixed disconnected correlator of a light quark disconnected loop 
at $am_q=0.01$ with a strange quark disconnected loop at mass $am_s=0.05$
in Fig.~\ref{fig:meffB676} gives a 
mass around 600 MeV. Unfortunately because this
number does not agree with that from the diagonal correlators
we do not use it to quote a physical mass for the 
lightest flavour singlet pseudoscalar meson.

\begin{figure}[t]
\resizebox{5.0in}{!}{\includegraphics{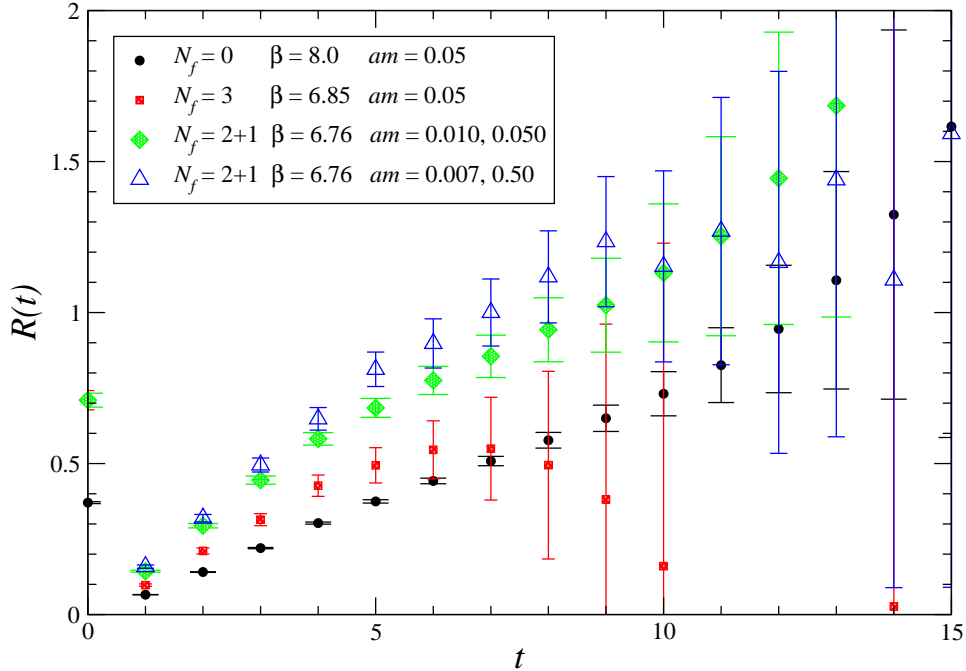}}
\caption{$D/C$ ratio for coarse MILC ensembles. For the $N_f=3$ and $N_f=2+1$
ensembles $R_{SU3}$ is displayed. For the quenched configurations the
single-flavour $R=D/C$ is plotted.
\label{allrat}}
\end{figure}

Fig.~\ref{allrat} shows the ratios of 
disconnected to connected correlators as defined by
(\ref{nondegenerate_ratio_eq}) in the non-degenerate unquenched case 
(the quenched data is also shown for comparison).
The ratio plots show that the errors get very
large as $t$ gets large. This makes the test that the 
$D(t)/C(t)$ tends to $1$ in the large $\Delta t$ limit difficult to judge
from looking at the graphs. With the unquenched data we constructed 
both $R(t)_{SU3}$ (defined by (\ref{nondegenerate_ratio_eq})), 
and the single-flavour $R(t)=D(t)/C(t)$ and 
fitted both to the model
\begin{equation}
R(t)= A_0 + B_0 e^{-(m_r t)}\ ,
\label{eq:DbyCration}
\end{equation}
where $m_r$ is the mass difference between the taste singlet pseudoscalar
meson and the connected pseudoscalar correlator.
The results for the three fit parameters are given in 
Table~\ref{tab:DbyCRESULTSB676}. In the table we compare $m_r$, 
obtained from the ratio, to
$a m_r^{\rm diff}$. We define the latter to be  the mass of the 
taste singlet pseudoscalar meson,
obtained from the matrix propagator (\ref{matrix_prop}) and 
listed in Table \ref{tab:massRESULTSB676DIS}, 
with the mass from the pure connected 
state (Table \ref{tab:massRESULTSB676CON}) subtracted off. 

Within the large 30\% errors
$A_0$ is 1 for the light quark as expected from theory. 
Also the values of $a m_r^{\rm diff}$ are consistent with 
$a m_r$. 
Given the noise in the calculation the starting
times in the fit could only be moved 1 or 2 time
units forwards or backwards from those quoted
in Table~\ref{tab:DbyCRESULTSB676}, so a more
convincing test would require much higher statistics.
Unfortunately we were unable to get stable fits 
of the $D/C$ ratio for the $N_f=3$ $\beta=6.85$  data
set, probably because
the statistics at $\beta=6.85$ were lower than for the 
other $\beta$ values.

\begin{table}[tb]
  \caption{
Parameters from the fit of (\ref{eq:DbyCration}) to the
unquenched data. The result for $m_r^{\rm diff}$ is from
the difference between the mass of the taste
singlet and the connected pseudoscalar channel.
The SU3 entry corresponds to the use of the
R ratio in eq.~\ref{nondegenerate_singlet_prop}.
}
\begin{center}
\begin{tabular}{|c|c|c|c|c|c|c|c|c|} \hline
$\beta$ & $m_q/m_s$ & Mass & region  & $A_0$ & $B_0$ & $a m_r$ &
$a m_r^{\rm diff}$ &
$\chi^2/dof$ \\ \hline
6.76 & 0.01/0.05 & 0.01 & 5 - 13   & 1.03(35) & -0.97(23) & 0.15(11) &
0.22(6) & 6.1/6  \\
6.76 & 0.01/0.05 & SU3 & 4 - 11   & 1.01(19) & -1.15(9) & 0.24(9) &
0.22(6) & 6.3/5  \\
6.76 & 0.007/0.05 & 0.007 & 4 - 9   & 1.14(39) & -1.47(23) & 0.24(18) &
0.31(9) & 1.5/3  \\
6.76 & 0.007/0.05 & SU3 & 4 - 8   & 1.15(30) & -2.09(91) & 0.35(24) &
0.31(9) & 1.2/2  \\
\hline
\end{tabular}
\end{center}
\label{tab:DbyCRESULTSB676}
\end{table}


\subsection{Topological Charge}

Smit and Vink \cite{Smit:1986fn} have pointed out that from the Atiyah-Singer
index theorem \cite{Atiyah:1968mp} it follows that the $\gamma_5\otimes{\bf 1}$
quark loop is a measure of the topological charge density. Having measured 
this quantity  as part of the disconnected correlator calculation 
we are in a position to 
compare the fermionic definition of topological charge
\begin{equation}
Q=m\kappa_p\langle {\rm Tr}\left(\gamma_5 M^{-1}\right)\rangle_U\ ,
\end{equation}
where $\kappa_p$ is a renormalisation factor, 
with the traditional gluonic definition
\begin{equation}
Q=\frac{g^2}{64\pi^2}\int dx\epsilon^{\mu\nu\rho\sigma}F^a_{\mu\nu}(x)F^a_{\rho\sigma}(x).
\end{equation}
Alles {\em et al.} have performed a similar comparison with Wilson fermions 
\cite{Alles:1998jq}. The MILC collaboration has extensively studied 
the topology on these lattices using gluonic definitions of the  
topological charge~\cite{Bernard:2003gq,Billeter:2004wx}.

We used the publicly available MILC code \cite{MILC_code} to measure
the topological charge.
The gauge fields were cooled by hypercubic 
blocking \cite{DeGrand:1997gu,Hasenfratz:2001hp,Bernard:2003gq} 
before the topological charge was measured.

\begin{figure}[tbh]
\resizebox{6.4in}{!}{\includegraphics{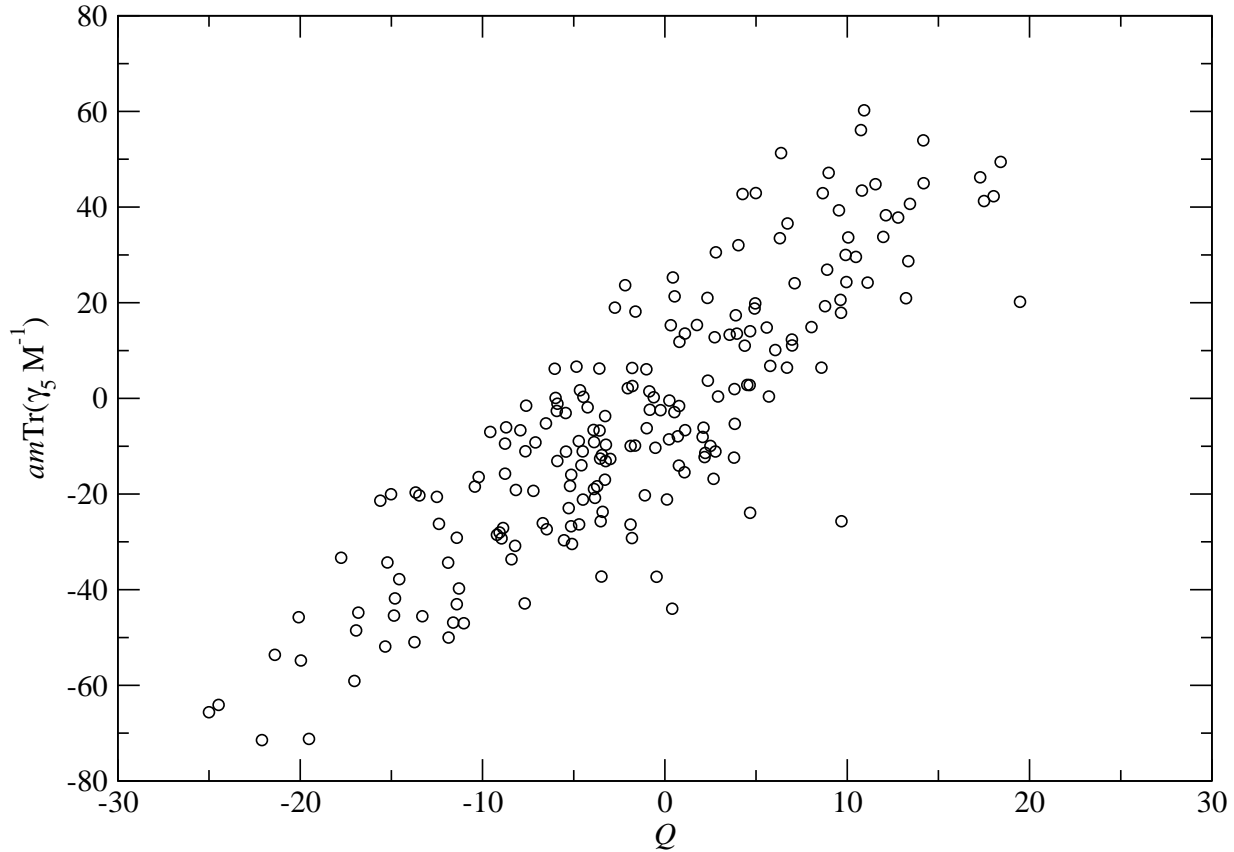}}
\caption{Scatterplot of fermionic and gluonic definitions of topological 
charge on $N_f=2+1$-flavour $\beta=6.76$ lattices. Light quark ($am=0.01$)
loops are used to calculate ${\rm Tr}\left(\gamma_5 M^{-1}\right)$.
\label{b676_topo_q_scatter}}
\end{figure}

A comparison of the two definitions of topological charge 
on 198 $N_f=2+1$, $\beta=6.76$ lattices is shown in 
Fig.~\ref{b676_topo_q_scatter}, and shows
that the gluonic definition of the topological charge 
is strongly correlated with the fermionic definition. This gives
us confidence that the staggered 
$\langle{\rm Tr}\left(\gamma_5 M^{-1}\right)\rangle $ operator,
which is a key building block in the $SP$ correlator, is behaving
as expected. A more empirical comparison between the fermionic and
gluonic topological charge requires knowledge of the 
renormalisation factors. 
Methods have been developed~\cite{Durr:2006ky,Alles:2006ur} to 
determine the matching renormalisation factor
between the two definitions of the topological charge,
based on their assumed equality.

\section{Statistics of Disconnected Correlators}\label{disc_stat_section}

Lattice QCD calculations that compute flavour singlet quantities
require many more gauge configurations than those that only
focus on flavour non-singlet quantities. For example, 
Fig.~\ref{binned_dc_rat_b800} shows that high statistics are
required even in quenched QCD to accurately compute the $D/C$ ratio. Although
recently there have been impressive algorithmic improvements 
in performing unquenched lattice QCD calculations, an increase
in the cost of an unquenched calculation by a factor of 10
is significant. This motivates a thorough study of the statistics
of disconnected correlators.

\begin{figure}[tbh]
\resizebox{6.4in}{!}{\includegraphics{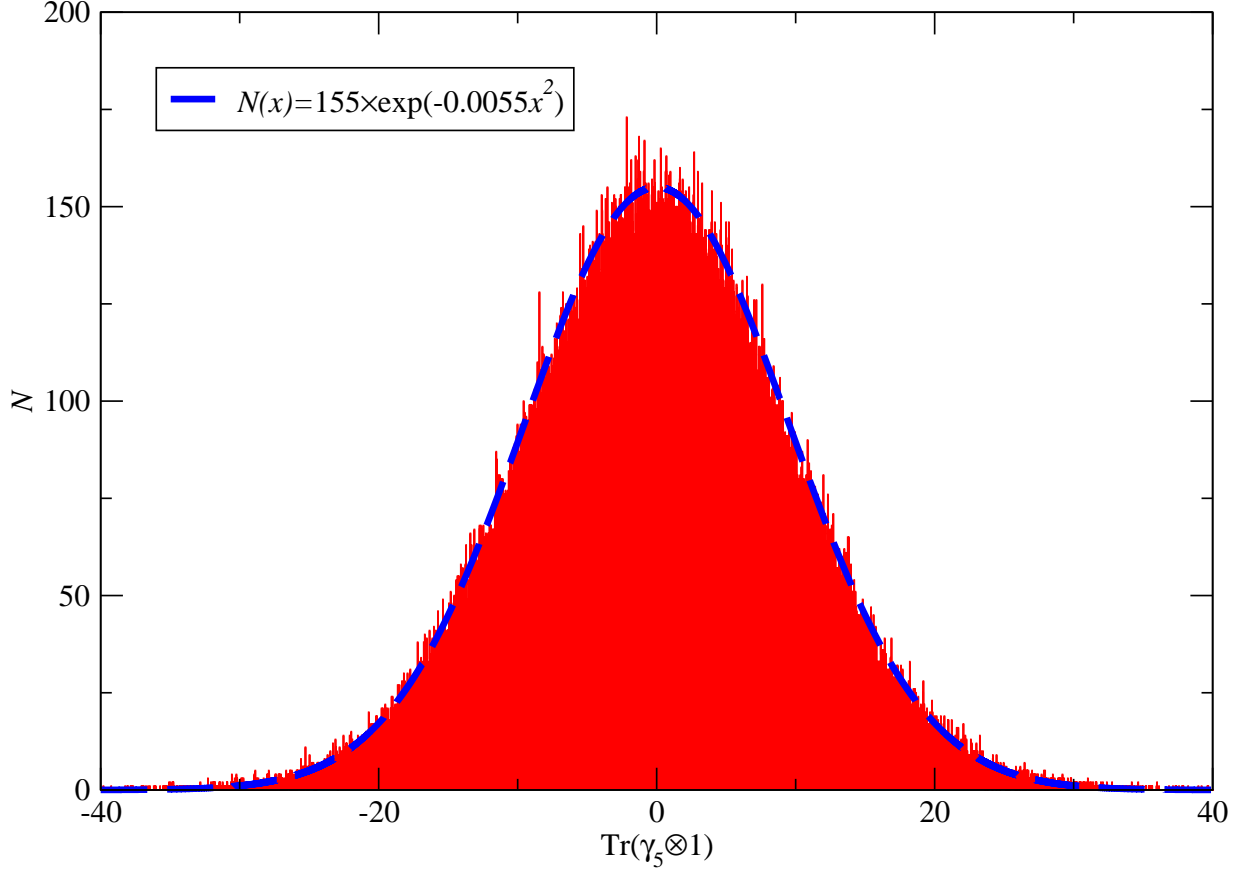}}
\caption{Histogram of 393856 ${\rm Tr}\gamma_5\otimes{\bf 1}$ measurements.
Analytic curve is an approximation rather than a fit.
\label{g5_loop_histo}}
\end{figure}

In Fig. \ref{g5_loop_histo}, we show a histogram  of 393856 
measurements (64 timeslices on 6154 quenched lattices) of the 
timeslice loop operator 
\begin{equation}
L(t)\equiv{\mathcal O}_{\gamma_5\otimes{\bf 1}}(t)
=\sum_{i\in {\rm timeslice}\hspace{1mm} t}\left[\Delta_{\gamma_5\otimes{\bf 1}}M^{-1}
\right]_{ii}\, .
\end{equation}

This is for the quenched data set at $\beta=8.00$ 
described in Table~\ref{extended_quenched_configs}
with valence quark mass $am=0.05$, and shows a clear Gaussian distribution for $L$.

\begin{figure}[tbh]
\resizebox{6.4in}{!}{\includegraphics{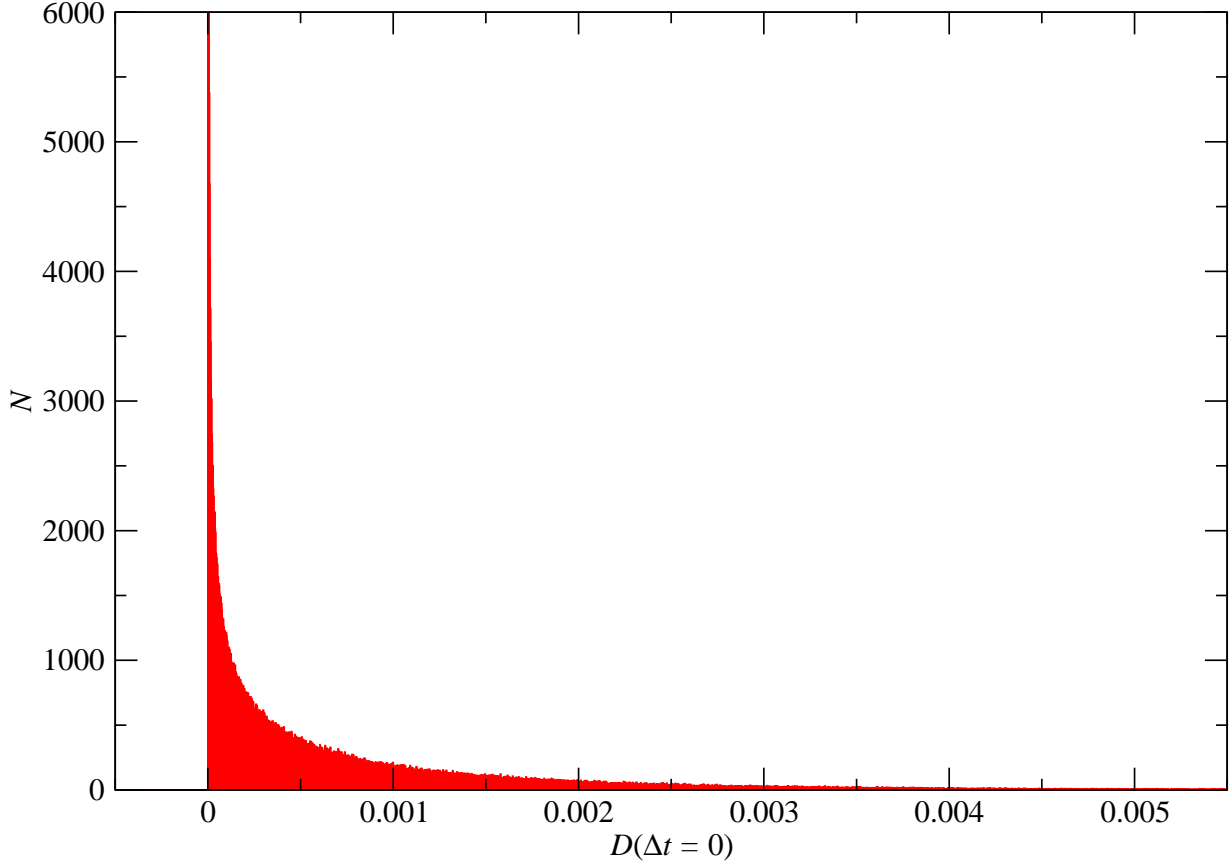}}
\caption{Histogram of  $D(\Delta t=0)$ measurements, 
and a curve generated with Equation \ref{dt00_distro}.
\label{g5_dcorr_histo_t0}}
\end{figure}

However the distribution of the disconnected correlators  $D(\Delta t)$ constructed
from these measurements is {\em not} Gaussian and standard error
estimation requires some care.
It is straightforward to describe the distribution of 
$D(\Delta t)$ in the limiting case of $\Delta t=0$.
In this case ($\Delta t=0$), $D(\Delta t)$ is just given by the square of 
the loop operator $L(t)$.
If  independent random variables $x_i$ (i.e. $L$)
are Gaussian distributed:
\begin{equation}
P(x)=A\exp{\left(-\frac{x^2}{\sigma_x^2}\right)},
\label{normal}
\end{equation}
then $w_i=x_i^2$ have a distribution
$Q(w)$:
\begin{equation}
Q(w)=P(x)\frac{dx}{dw} = \frac{A}{\sqrt{w}}\exp{\left(-\frac{w}{\sigma_x^2}\right)}.
\label{dt00_distro}
\end{equation}
Note that the domain of $Q(w)$ is $w\geq 0$. This is consistent with 
the observed distribution of quenched disconnected correlator measurements $D$ at
$\Delta t=0$ in Fig. \ref{g5_dcorr_histo_t0}.

\begin{figure}[tbh]
\resizebox{6.4in}{!}{\includegraphics{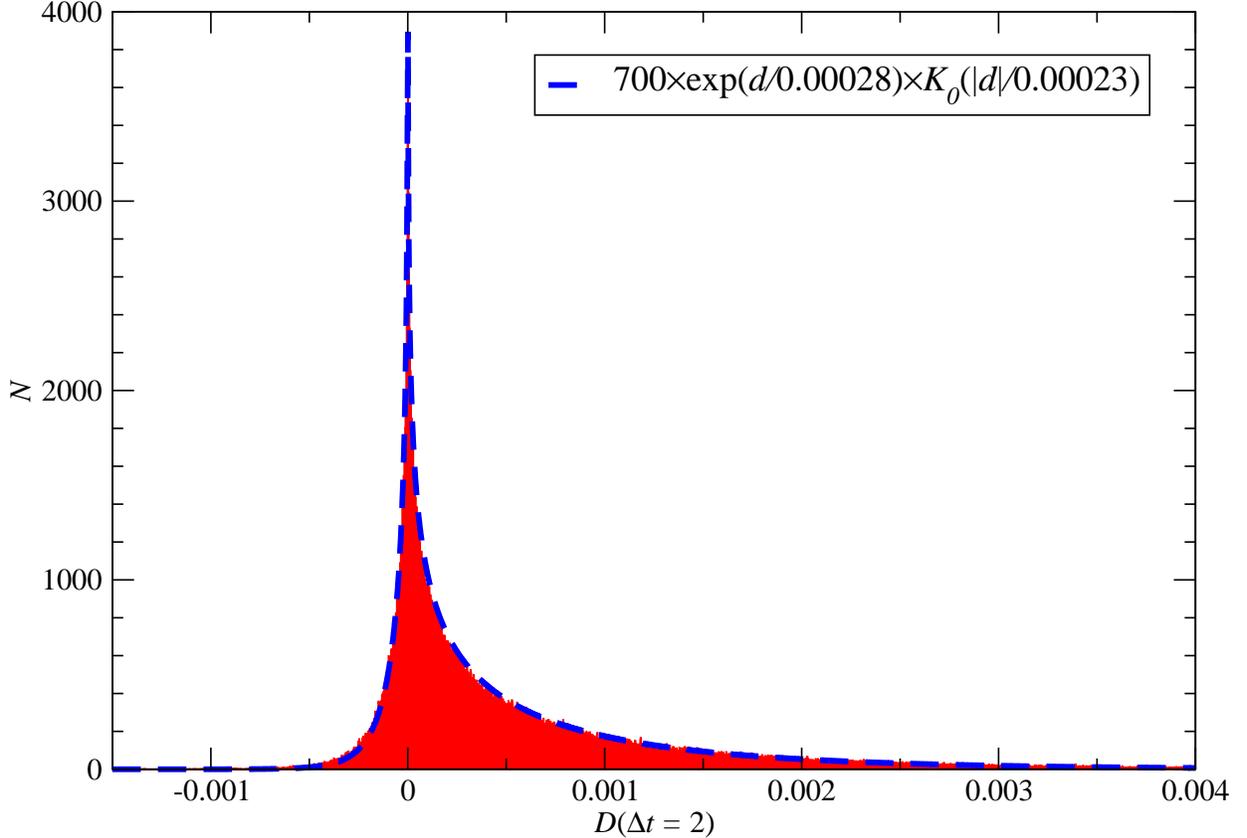}}
\caption{Histogram of 393856 $D(\Delta t=2)$ measurements on 6154 quenched 
lattices, and a curve generated with Equation \ref{dcorr_dist_form}.
\label{g5_dcorr_histo_t2}}
\end{figure}

In the region of interest, $0 < \Delta t < \Delta t_{\rm max}$ (with $\Delta t_{\rm max}\approx 15$),
$L(t)$ and $L(t+\Delta t)$ are
correlated non-trivially and the mean $\bar{D}(\Delta t)>0$. 
In Fig. \ref{g5_dcorr_histo_t2}, the distribution of
$D(\Delta t=2)$ is clearly seen to have asymmetric tails.

Consider two Gaussian-correlated variables $x$ and $y$ with the same 
variance $\sigma^2$. They have the joint probability distribution
\begin{equation}
P(x,y)=\kappa\exp\left[-\left(Ax^2 + A y^2 + 2Bxy\right)\right].
\label{xy_corr_dist}
\end{equation}
Changing variables to $z_i=x_iy_i$ and $\theta_i=\arctan(y_i/x_i)$ we get the joint 
probability for $z$ and $\theta$:
\begin{eqnarray}
\tilde{Q}(z,\theta)&=&P(x,y)\left|\frac{\partial x\partial y}{\partial z\partial\theta}\right|\nonumber\\
&=&\kappa\exp\left[-\left(\frac{2Az}{\sin(2\theta)}+2Bz\right)\right]\left|\frac{1}{\sin 2\theta}\right|.
\end{eqnarray}
Integrating out $\theta$, treating positive and negative $z$ regions
separately (for each it suffices to integrate from zero to $\pm\pi/4$ 
and multiply by 4), we get
\begin{eqnarray}
Q(z)&\equiv&2\int^{\pi/4}_0d\theta\tilde{Q}(z,\theta)\nonumber\\
&=&8\kappa \exp\left(-2Bz\right) \int^{\pi/4}_0d\theta 
\exp\left[-\left(\frac{2Az}{\sin(2\theta)}\right)\right]
\left(\frac{1}{\sin(2\theta)}\right)\nonumber\\
&=& 4\kappa \exp\left(-2Bz\right)K_0\left(2A\left|z\right|\right),
\label{dcorr_dist_form}
\end{eqnarray}
where $K_0(z)$ is a modified Bessel function of the second kind. This agrees 
nicely with, for example, $D(\Delta t=2)$ measurements for the 
quenched $\beta=8.00$, $am=0.05$ ensemble data, shown in Fig. \ref{g5_dcorr_histo_t2}.

\begin{figure}[tbh]
\resizebox{6.4in}{!}{\includegraphics{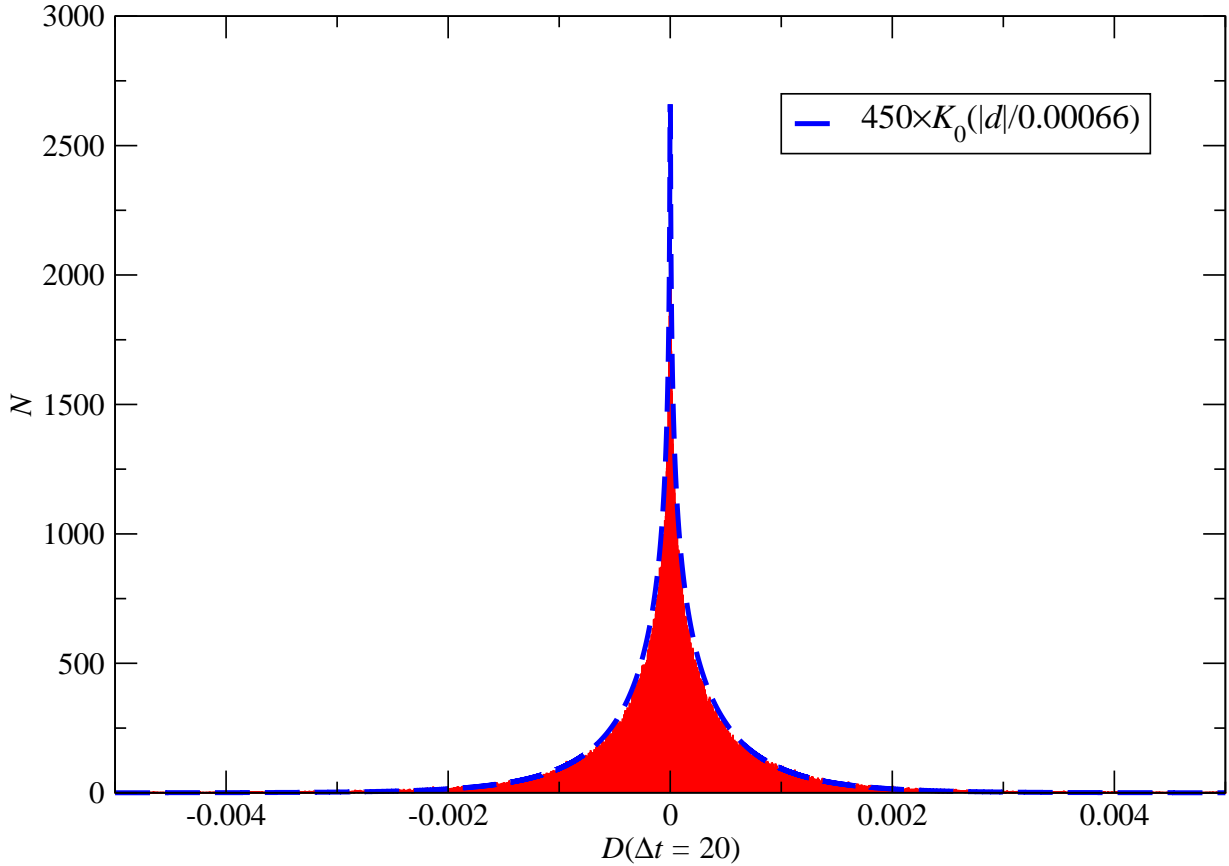}}
\caption{Histogram of  $D(\Delta t=20)$ measurements, and a plot of a curve
generated from Equation \ref{uncorr_gauss_product}, the theoretical
distribution of the product of two completely uncorrelated Gaussian variables.
\label{g5_dcorr_histo_t20}}
\end{figure}

We see that in the case of $\Delta t \rightarrow {\rm large}$, 
$D(\Delta t)\rightarrow 0$ and  $B\rightarrow 0$. Equation (\ref{xy_corr_dist}) then
factorises into separate $x$ and $y$ parts and
\begin{equation}
\tilde{Q}(z,\theta) = P(x)P(y)\left|\frac{\partial x\partial y}{\partial z\partial \theta}\right|.
\end{equation}
As expected, we get 
\begin{equation}
Q(z)=4\kappa K_0\left(2A\left|z\right|\right).
\label{uncorr_gauss_product}
\end{equation}
This is 
consistent with the measured quenched disconnected correlators at large time 
separation.  Fig.~\ref{g5_dcorr_histo_t20} shows these
correlators for $\Delta t=20$ together with a plot of the modified 
Bessel function form (\ref{uncorr_gauss_product}).

In all cases the most likely measurement is $D(\Delta t)=0$, irrespective of
the mean.
Clearly the signal we are trying to resolve in taking the mean over
disconnected correlator measurements comes from the asymmetry in the 
distribution, which is induced by the exponential factor in 
(\ref{dcorr_dist_form}). When the mean disconnected correlator is small
the distribution is almost symmetric and a large proportion of the
signal comes from the tails of the distribution. The number of data points in
the tails of the distribution many standard deviations from the mean is far greater 
than in a Gaussian distribution (Table \ref{distributions}) 
and act as a great lever arm on 
the mean itself. When $\Delta t$ is large, both  $D(\Delta t)$ and 
$C(\Delta t)$ are small and consequently the $D/C$ ratio experiences large
fluctuations.

\begin{figure}[tbh]
\resizebox{6.4in}{!}{\includegraphics{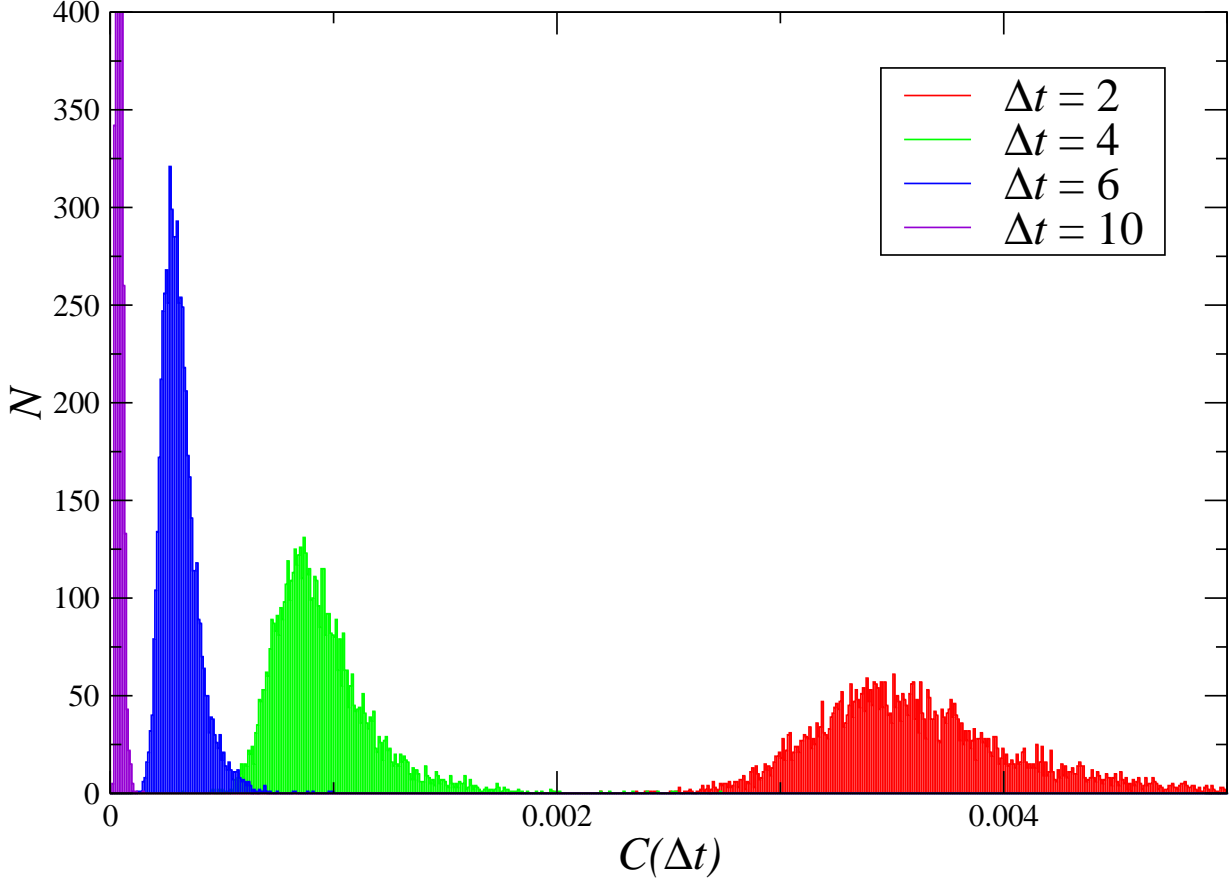}}
\caption{Histogram of connected singlet pseudoscalar $(\gamma_5\otimes{\bf 1})$
 measurements with $am=0.05$ on 6154 $\beta=8.00$ quenched lattices.
\label{g5_ccorr_histo_t0-10}}
\end{figure}

All of this is in marked contrast to the statistics of the connected 
correlator, the histograms of which are less peaked and have less 
pronounced tails (Fig. \ref{g5_ccorr_histo_t0-10}).
Another difference is obvious --- up to differences in the asymmetry, the 
width of the distribution of disconnected correlator measurements remains 
approximately constant with respect to $\Delta t$, whereas the connected correlator 
distributions are of similar shape with the width roughly proportional to the 
mean.

The non-Gaussian nature of the distribution of disconnected correlator 
measurements has significant consequences for the interpretation of error
estimates on these correlators, and upon derived quantities. We use bootstrap sampling
methods to estimate the error on the mean of quantities of interest. Although
the distribution of $D$ is non-Gaussian, the central limit theorem ensures
that the distribution of the mean $\bar{D}$ itself is Gaussian. Given large
enough samples we should therefore be able to safely use the 
bootstrap resampling method to study the error on the mean using the usual confidence interval methods, but caution is called for with smaller samples.

\begin{figure}[tbh]
\resizebox{6.4in}{!}{\includegraphics{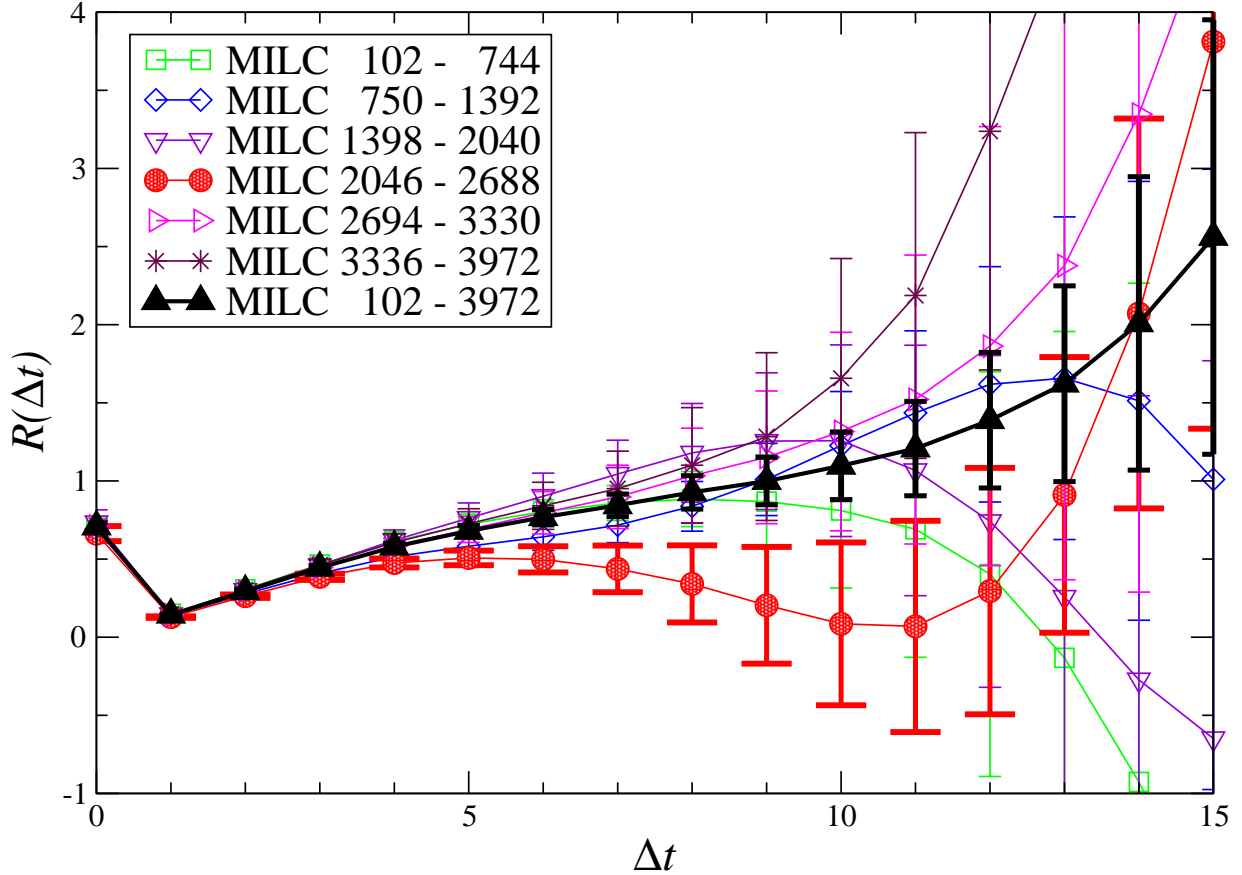}}
\caption{Binned D/C ratio for $\beta=6.76$ $am=0.01,0.05$. \label{binned_dc_rat_b676}}
\end{figure}

\begin{figure}[tbh]
\resizebox{6.4in}{!}{\includegraphics{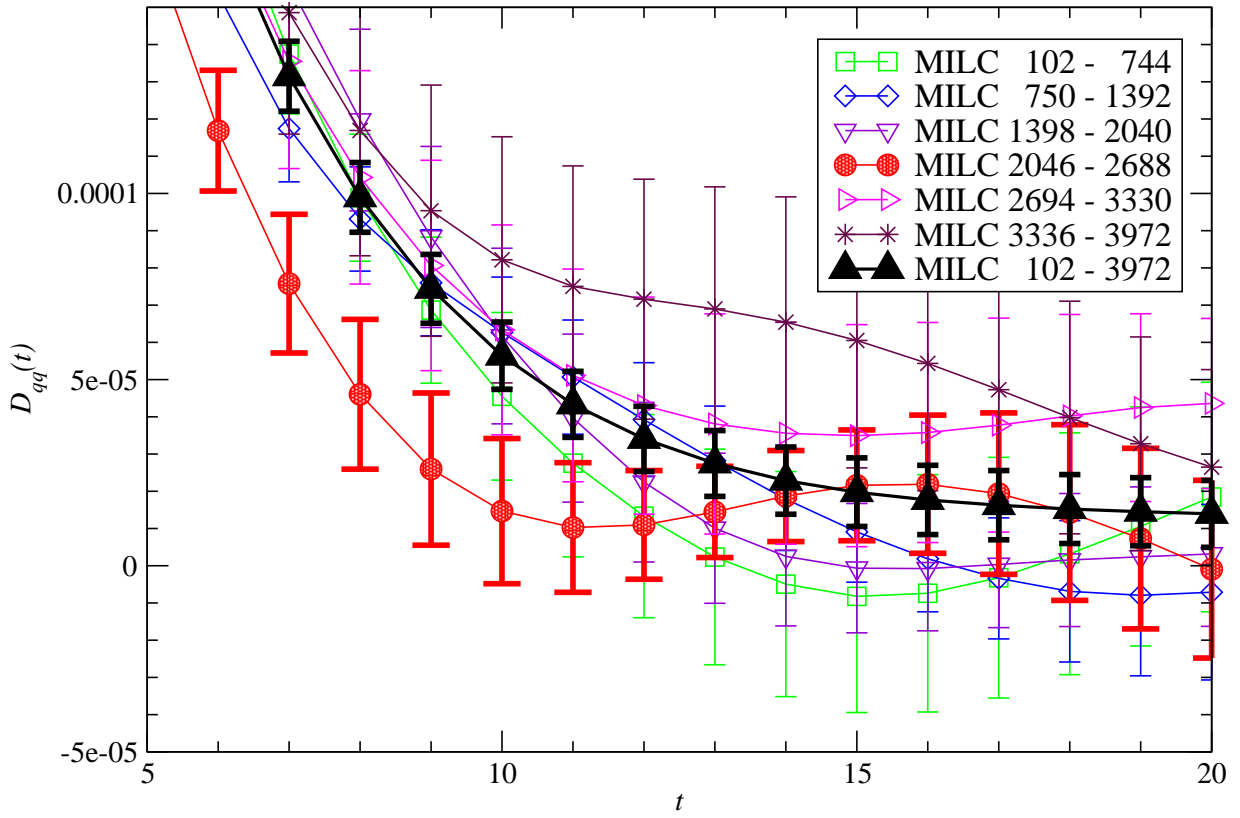}}
\caption{Binned $D_{qq}$ correlators for $\beta=6.76$ $am=0.01$. \label{binned_dcorr_b676}}
\end{figure}

As an example we consider the 658 configurations of the MILC coarse ensemble 
with $\beta=6.76$ and sea-quark masses $am=0.01$ and $0.05$. 
We measured the disconnected correlators and 
calculated the $D/C$ ratio $R(\Delta t)$ as described in
Section  \ref{unquence_sec}, 
We dropped  12 configurations with trajectory index less than 100 and separated the 
remaining 646 into six bins of about 108 configurations each. 
Fig. \ref{binned_dc_rat_b676} shows the  $D/C$ ratios for the subsets so defined, where
the errors were calculated by bootstrap
using the usual $1\sigma$ ($\sim68\%$ confidence level) definition.  
We notice 
that, at the modest time-separation of $\Delta t = 10$, two subsets (the first 
and last) are nearly a standard deviation from the value obtained with all the 
data, and one subset (the third, highlighted with a filled circle symbol) is 
nearly two standard deviations from the value obtained with all 646 
configurations.
We can immediately trace this to the disconnected correlators (in Fig. \ref{binned_dcorr_b676}, the subsets show exactly the same discrepancy in the light-quark disconnected correlator $D_{qq}$). 

\begin{figure}[tbh]
\resizebox{6.4in}{!}{\includegraphics{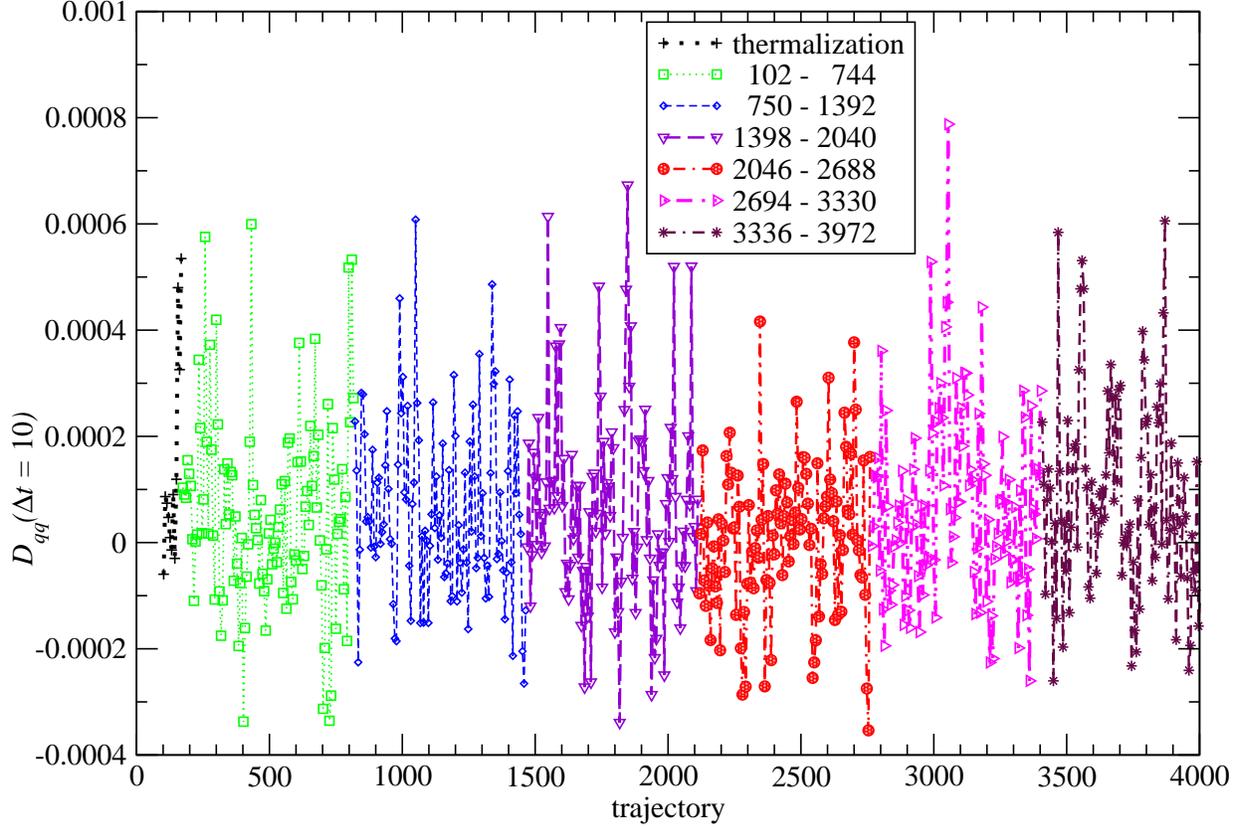}}
\caption{Timeseries of $D_qq$ correlator measurements (averaged over 
timeslices) for $\beta=6.76$ $am=0.01$. \label{dcorr_tseries_b676}}
\end{figure}

An underestimate of autocorrelation times can cause underestimation of
errors, leading to an apparent disagreement betwwen subsets of 
measurements. 
In \cite{Bernard:2003gq} the MILC collaboration report long autocorrelation
times for some $28^3\times 96$, $a\sim 0.1$fm ensembles, but not in the
`coarse' $20^3\times 64$ ensembles. Like MILC, we found no measurable
autocorrelation of the
topological charge in the coarse $\beta=6.76$, $am=0.01,0.05$ ensemble. The related,
but more relevant quantity, the pseudoscalar singlet disconnected 
correlator
is also suitably decorrelated.   For $D_{qq}(\Delta t = 10)$ we get
$\tau_{\rm ac}=1.09(14)$. The corresponding  timeseries is shown in
Fig.~\ref{dcorr_tseries_b676}.

\begin{figure}[tbh]
\resizebox{6.4in}{!}{\includegraphics{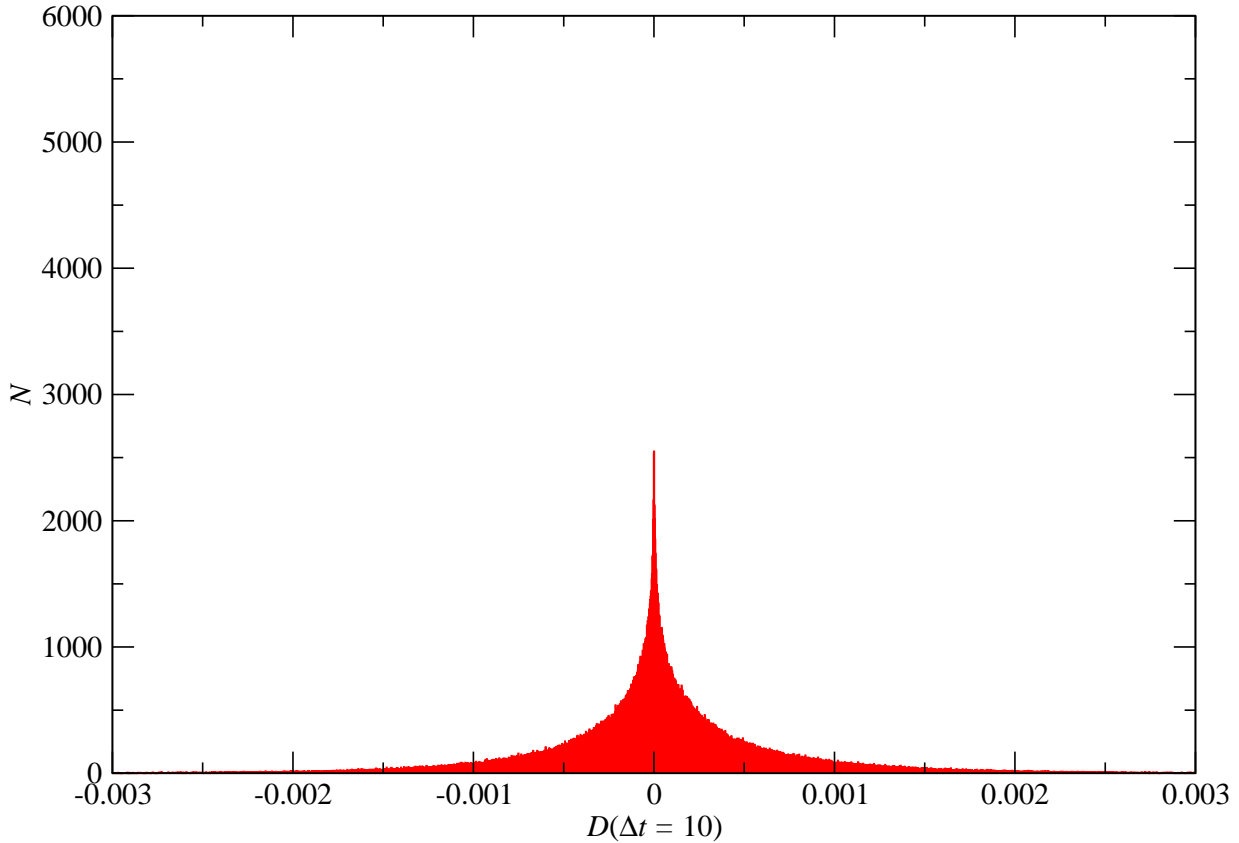}}
\caption{Histogram of 393856  $D(\Delta t=10)$ measurements on 6154 quenched 
lattices.
\label{g5_dcorr_histo_t10}}
\end{figure}

So the autocorrelation time is not the 
cuplrit here. It is clear from inspection that subsets of the timeseries where 
$D_{qq}(\Delta t)$ is particularly deviant from the mean correspond to those
with a relative abundance (or deficit) of spikes or data points falling in the
tails of distributions such as in Fig. \ref{g5_dcorr_histo_t10}.\footnote{The points 
in the timeseries of $D_{qq}(\Delta t)$ measurements are of course the average 
of $N_t=64$ correlated points from a distribution like 
Fig. \ref{g5_dcorr_histo_t10}, so the tails and peak are somewhat less 
severe.}

Quenched ensembles exhibit the same features and this is particularly 
clear from Fig.~\ref{binned_dc_rat_b800}.
We now address the validity of the error bars on the 
ratio plot for the 400-configuration subsamples.
Fig. \ref{binned_dc_rat_b800}
shows the ratio $R=D/C$ for the first 400 configurations of each stream.
The standard error for each set is computed by bootstrap, with a bin size of ten
configurations. Bootstrap and jackknife give consistent error estimates.
From our full 6154-configuration quenched ensemble we can make 15 bins 
of 400 configurations (four of the S0 bins are not shown in 
Fig. \ref{binned_dc_rat_b800}).  Of these, eight have a value of 
$ R(10)$ within its own standard error of 
$\overline{R(10)}=0.28$ (the value for all 6154 configurations). Three 
give a value between one- and two- sigma away, and five have a value
between two- and three- sigma away. Assuming the means are normally distributed
(as discussed above) we would
expect that only one of the fifteen bins would 
give a value more than $2\sigma$ from the ``true'' value. Although fifteen is 
a small number of bins, this distribution leads us to consider that the 
size of the error on these bins may be slightly underestimated.

If (\ref{xy_corr_dist}) is indeed the form of the distribution of loop 
operator measurements, then the above suggests that fitting 
the histogram of disconnected correlator measurements to 
(\ref{dcorr_dist_form}) may be a useful method of extracting 
$D(\Delta t)= \langle z \rangle$. From (\ref{xy_corr_dist}) we 
easily calculate that
\begin{equation}
\sigma^2_x=\sigma^2_y\equiv\langle x^2\rangle=\frac{A}{A^2-B^2}
\label{x_variance}
\end{equation}
and
\begin{equation}
\langle xy \rangle = \frac{1}{2}\frac{-B}{A^2-B^2}.
\label{xy_correlation}
\end{equation}
From the Gaussian histogram of loop 
operator measurements one can extract the overall normalization $\kappa$,
and the width of the Gaussian puts a constraint on $A$ and $B$ in the form of 
(\ref{x_variance}), such that there remains effectively one parameter to fit
to extract $\langle z \rangle$ with (\ref{xy_correlation}). 
The advantage may be that such a fit would 
depend less on the fluctuating tails of the distribution than the
average does. We leave evaluation of this method for future work.

\begin{table}
\begin{tabular}{@{}lcccc}
\hline
Distribution:& Normal & $P[D(\Delta t = 10)]$ & $K_0(z)$ & $P[D(\Delta t = 20)]$ \\
\hline
$1\sigma $ &0.317311     & 0.210280  & 0.208994 & 0.208147 \\
$2\sigma $ &0.045500     & 0.061705  & 0.061829 & 0.061289 \\
$3\sigma $ &0.002700     & 0.019172  & 0.019639 & 0.019606 \\
$4\sigma $ &0.000063     & 0.006297  & 0.006460 & 0.006751 \\
$5\sigma $ &5.73303e-07  & 0.002232  & 0.002170 & 0.002272 \\
$6\sigma $ &1.97318e-09  & 0.000779  & 0.000740 & 0.000779 \\
$7\sigma $ &2.55962e-12  & 0.000322  & 0.000255 & 0.000297 \\
$8\sigma $ &1.22125e-15  & 0.000145  & 0.000088 & 0.000104 \\ 
$9\sigma $ &             & 0.000046  & 0.000031 & 0.000038 \\
$10\sigma $ &            & 0.000025  & 0.000011 & 0.000025 \\
\hline
\end{tabular}
\caption{\label{distributions} The integrated fraction of data occurring within
a radius of multiples of the standard deviation $\sigma$ from the mean. 
Columns two and four are analytic normal (Eq. \ref{normal}) and 
modified Bessel (Eq. \ref{uncorr_gauss_product}) distributions,
respectively. Columns three and five are the observed distribution of $D(\Delta t=10)$ and $D(\Delta t=10)$ for 393856 measurements on quenched lattices.}
\end{table}


\section{Number of configurations needed} \label{se:numberneeded}

Flavour singlet lattice spectroscopy is difficult mainly because of the 
difficulty in getting precise determinations of the pseudoscalar disconnected 
correlators --- disconnected correlators are far more sensitive to fluctuations in
the sea than connected correlators. Furthermore the pseudoscalar quark loop 
operator is a measure of the topological charge which is known to be plagued 
with long autocorrelation times \cite{Bernard:2003gq}.

We treat this work as a first step toward a more systematic exploration of 
pseudoscalar flavour singlets with staggered quarks. Results such as those 
depicted in Figures \ref{binned_dc_rat_b676} and \ref{binned_dc_rat_b800}              
and the difficulty in fitting 
indicate that far more configurations are necessary. 

On average, over an ensemble, the error on the $\gamma_5\otimes {\bf 1}$
loop operator is independent of timeslice $t$. A consequence of this is that
the product
\begin{equation}
{\mathcal O}_{\gamma_5\otimes{\bf 1}}(t){\mathcal O}_{\gamma_5\otimes{\bf 1}}(t+\Delta t),
\end{equation}
and hence the disconnected correlator $D(t)$, has an error that is roughly
constant in $t$ while the size of the correlator decreases over many orders of 
magnitude as $t$ increases. See, for example,
Fig. \ref{all_Dcorrs_b676m010m050}.

\begin{figure}[tbh]
\resizebox{5.0in}{!}{\includegraphics{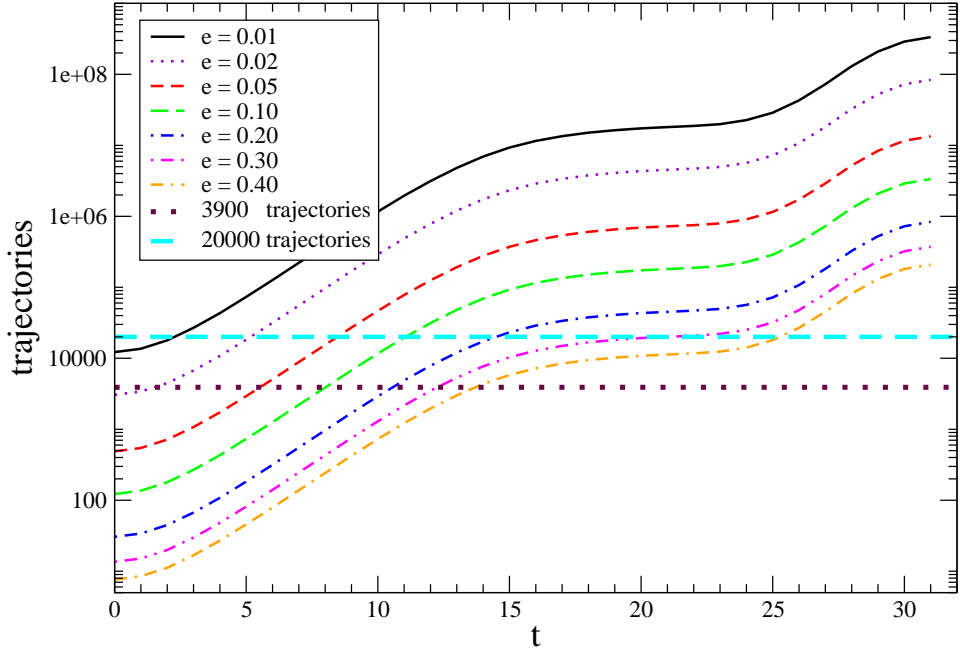}}
\caption{Estimated trajectories needed for improved precision of $D_{qq}(t)$ 
for $\beta=6.76$ $am=0.01$ $20^3\times 64$ lattices, where $e(t)\equiv \overline{\sigma_{D_{qq}}}/D_{qq}(t)$. 
\label{traj_for_precision}}
\end{figure}

Since the gauge contribution to the statistical error of the disconnected
correlator scales as $1/\sqrt{N_{\rm cfg}}$, it is straightforward to estimate
the number of configurations needed to resolve the light-light disconnected
correlator (for example) out to some time-separation $t$ to a given precision.
We use the $\beta=6.76$, $am=0.01, 0.05$ MILC ensemble, which has 658 
configuration separated by 6 trajectories. We have further grouped the lattices
into bins of 10 lattices (thus eliminating 8). So with 3900 trajectories
the gauge error on $D_{qq}(t)$ with our normalization is roughly $10^{-5}$, 
independent  of $t$ (see Fig. \ref{gauss_Ns_dependence_kcp}.) In Fig.
\ref{traj_for_precision} we use the obtained values of $D_{qq}(t)$ to estimate
 the number of trajectories needed for various values of precision $e$ as a function of time
separation. The dotted horizontal line represents the current 3900
trajectories, showing that we currently have 20\% resolution only out to 
$t=10$ and 10\% resolution out to $t=8$. 

It is feasible with modern computational resources to produce ensembles using 
similar parameters with $\sim 2\times 10^4$ trajectories.
The position of the dashed horizontal line 
represents the precision one hopes to derive
from these extended statistics. We hope that $D_{qq}$ would be resolvable to 
20\% precision to $t=15$ and to 10\% precision to $t=11$.
The use of anisotropic lattices may be useful
in allowing more time slices where the signal
is bigger than the noise~\cite{Morrin:2006tf,Levkova:2006gn}.
However, the tuning of anisotropic lattice actions is non-trivial
for unquenched lattice QCD calculations.

The issue of autocorrelation times has been found not to 
be a problem at the current
lattice spacing. In the event of future work on finer lattices --- a 
necessary step for controling taste-breaking effects --- care should be 
taken. However, the MILC collaboration 
report in \cite{Bernard:2003gq} that topological 
charge autocorrelation times grow with fermion mass, so targeting lighter 
quark masses should help ease the problem.
We are concerned with correlations of the density of
${\rm Tr}\Delta_{\gamma_5\otimes{\bf 1}}M^{-1}$ between timeslices.
The success of MILC's method of subdividing large lattices for dealing 
with slow topological modes suggests that autocorrelation of the 
topological charge density on timeslices is likely to be less of a 
problem than that of total topological charge.

\begin{figure}[tbh]
\resizebox{5.0in}{!}{\includegraphics{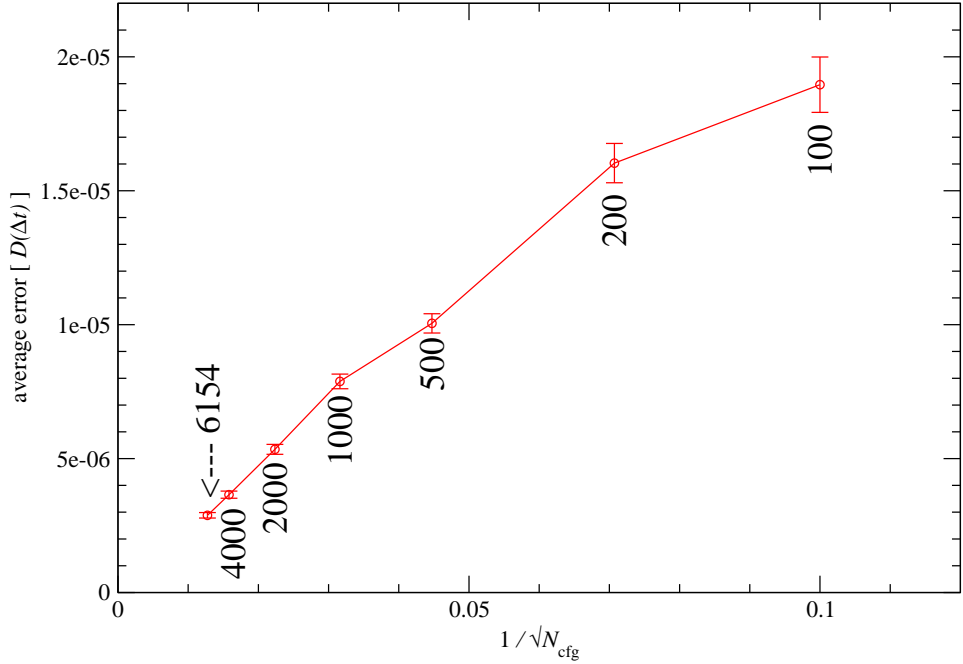}}
\caption{Error on $D(\Delta t)$ , averaged over $\Delta t$, for $\beta=8.00$ 
quenched configurations with $am=0.05$, as a function of 
$1/\sqrt{N_{\rm cfg}}$.\label{err_vs_Ncfg}}
\end{figure}

\section{Conclusions}  \label{se:conclusions}

We have investigated a number of different algorithms
to compute the disconnected diagrams required for the
correlators for the singlet pseudoscalar mesons. We
found that the algorithm proposed by
Venkataraman and Kilcup~\cite{Venkataraman:1997xi}
was the most efficient for our purposes.

Our results for the ratio of disconnected to
connected diagrams (Fig. \ref{allrat})
do not really show a convincing difference between
the quenched (\ref{quenched_ratio_eq})
and unquenched (\ref{degenerate_ratio_eq}) theories. At large time 
separations
the error on the ratio becomes large and it is
not clear that it asymptotes at 1. This is in
contrast to the work by
Venkataraman and Kilcup~\cite{Venkataraman:1997xi}
where the difference
between quenched QCD and unquenched QCD for the
ratio was clear.
It is clear that higher statistics are needed for
a definite conclusion on the large-time behaviour
of the ratio of the disconnected to connected
correlator, as well as to extract reliable
masses.

We have started to study the effect of the mixing between light and
strange interpolating operators. This is an essential part of the
physics of the $\eta$ and $\eta^\prime$ meson. 
The older unquenched calculations using the clover
and Wilson fermion action~\cite{Allton:2001sk},
that were done with quark masses that were
heavier than half the strange quark mass,
produced results consistent with quenched QCD
calculations for the majority of quantities.
This suggested that the
strange quark would not play a significant role in the dynamics
of $2+1$-flavour calculations. However the strange quark clearly
plays an important role in the physics of the $\eta$
and $\eta'$ mesons.

The correlators of the singlet pseudoscalar meson
are closely related to eigenvalues of the
quark operator~\cite{Neff:2001zr}.
This calculation has been done at a lattice spacing
of 0.12 fm. The calculation of the eigenvalues of the
Asqtad improved staggered fermion operator
in quenched QCD at a lattice spacing of 0.09fm,
by Follana at al.~\cite{Follana:2005km},
do not show convincing clustering
of the eigenvalues into quartets. However, the clustering of the eigenvalues
for the HISQ improved staggered action is convincing.
This is some evidence that the computation
of the spectroscopy of flavour singlet pseudoscalar mesons
may require gauge configurations with finer
lattice spacing than used here.

Although this lattice calculation has not produced
the spectacular agreement with experiment that
other parts of the improved staggered program
have achieved~\cite{Davies:2003ik,Aubin:2004wf}, 
we have not yet seen any ``show stoppers".
We are currently generating additional configurations
to increase the
statistics and in order to go to a finer lattice spacing.

\section*{ACKNOWLEDGEMENTS}

We are grateful to the ULgrid project of the
University of Liverpool for computer time.
We thank Chris Michael, Christine Davies, Claude Bernard, 
Greg Kilcup, Steve Sharpe, Zbyszek Sroczynski, Steve Miller, 
and Tommy Burch for discussions.

We thank Robert Edwards and Balint Joo for
help with Chroma~\cite{Edwards:2004sx}. 
We thank the MILC collaboration for making 
available their gauge configurations.
This work was in part based on the MILC collaboration's 
public lattice gauge theory code (see http://physics.utah.edu/$\sim$detar/milc.html)

\bibliographystyle{h-physrev2}
\bibliography{eta}

\end{document}